\documentclass[letterpaper,twocolumn,10pt]{article}
\usepackage[table]{xcolor}
\usepackage{usenix}
\newif\ifAnon\Anonfalse

\usepackage[utf8]{inputenc} 
\usepackage[T1]{fontenc}    
\usepackage{graphicx}       
\usepackage{amsmath}        
\usepackage{amssymb}
\usepackage{amsthm}
\usepackage{pifont}

\usepackage{bm}            
\usepackage{array}          
\usepackage{hyperref}       
\usepackage[english]{babel} 
\usepackage{enumitem}       
\usepackage{algorithm}
\usepackage{algpseudocodex}
\usepackage{booktabs}
\usepackage{tikz}
\usepackage{xspace}
\usepackage{cleveref}
\crefname{ALC@line}{line}{lines}
\Crefname{ALC@line}{Line}{Lines}
\usepackage{pgfplots}
\usepackage{mathptmx}
\usepackage{adjustbox}
\usepackage[clock]{ifsym}
\usepackage[normalem]{ulem}
\pgfplotsset{compat=1.18}          
\usepgfplotslibrary{statistics}    
\usepackage{tcolorbox}
\usepackage{stfloats}
\usetikzlibrary{decorations.pathreplacing,chains} 
\usetikzlibrary{positioning,overlay-beamer-styles,arrows.meta,matrix, calc, backgrounds}

\usepackage{tabularx}
\usepackage{booktabs}
\usepackage{array}
\usepackage{threeparttable}
\usepackage{oplotsymbl}
\usepackage[table]{xcolor}
\definecolor{rowgray}{gray}{0.95}
\newcolumntype{L}[1]{>{\raggedright\let\newline\\\arraybackslash\hspace{0pt}}m{#1}}
\newcolumntype{C}[1]{>{\centering\let\newline\\\arraybackslash\hspace{0pt}}m{#1}}
\newcolumntype{R}[1]{>{\raggedleft\let\newline\\\arraybackslash\hspace{0pt}}m{#1}}
\newcolumntype{P}[1]{>{\raggedright\arraybackslash}p{#1}}
\newcommand{\cmarkfull}{\multicolumn{1}{c} {\circletfill \phantom{*}}}
\newcommand{\cmarkempty}{\multicolumn{1}{c} {\circlet \phantom{*}}}
\newcommand{\cmarkhalf}{\multicolumn{1}{c} {\circletfillhl \phantom{*}}}

\newlist{titemize}{itemize}{1}
\setlist[titemize,1]{topsep=10pt, partopsep=0pt, nosep, left=0pt, label=\textbullet}

\newcommand{\etal}{et~al.\xspace}
\newcommand{\ie}{\emph{i.e.},\xspace}
\newcommand{\eg}{e.g.,\ }
\newcommand{\knocknock}{Knock-Knock\xspace}
\newcommand{\gf}{$\operatorname{GF}(2)$\xspace}
\newcommand{\FlushReload}{Flush+\allowbreak Reload\xspace}
\newcounter{observation}

\newcommand{\scalar}[1]{#1}
\newcommand{\addr}[1]{\uppercase{#1}}
\newcommand{\mymatrix}[1]{\uppercase{\boldsymbol{#1}}}
\newcommand{\myset}[1]{\mathbb{\uppercase{#1}}}
\newcommand{\card}[1]{|#1|}

\newcommand{\summarybox}[1]{%
  \refstepcounter{observation}%
  \begin{tcolorbox}[colback=gray!10]
    \textbf{Summary~\theobservation:} #1
  \end{tcolorbox}%
}

\newcommand{\highlightbox}[1]{%
  \begin{tcolorbox}[colback=gray!10]
    #1
  \end{tcolorbox}%
}

\newcommand{\todo}[1]{}
\renewcommand{\todo}[1]{{\color{red}{#1}}}

\begin{document}
\title{Knock-Knock: Black-Box, Platform-Agnostic DRAM Address-Mapping Reverse Engineering}

\ifAnon
\author{
{\rm Anonymous Authors}
}

\else
\author{
{\rm Antoine Plin}\\
GeorgiaTech, ESILV, CentraleSupelec, Inria, CNRS, IRISA
\and
{\rm Lorenzo Casalino}\\
CentraleSupelec, Inria, CNRS, IRISA
\and
{\rm Thomas Rokicki}\\
CentraleSupelec, Inria, CNRS, IRISA
\and
{\rm Ruben Salvador}\\
CentraleSupelec, Inria, CNRS, IRISA
}
\fi

\maketitle

\begin{abstract}
Modern Systems-on-Chip (SoCs) employ undocumented linear address-scrambling functions to obfuscate DRAM addressing, which complicates DRAM-aware performance optimizations and hinders proactive security analysis of DRAM-based attacks; most notably, Rowhammer.
Although previous work tackled the issue of reversing physical-to-DRAM mapping, existing heuristic-based reverse-engineering approaches are partial, costly, and impractical for comprehensive recovery.
This paper establishes a rigorous theoretical foundation and provides efficient practical algorithms for \emph{black-box, complete physical-to-DRAM address-mapping recovery.}

We first formulate the reverse-engineering problem within a linear algebraic model over the finite field \gf.
We characterize the timing fingerprints of row-buffer conflicts, proving a relationship between a bank addressing matrix and an empirically constructed matrix of physical addresses.
Based on this characterization, we develop an efficient, noise-robust, and fully platform-agnostic algorithm to recover the full bank-mask basis in polynomial time, a significant improvement over the exponential search from previous works.
We further generalize our model to complex row mappings, introducing new hardware-based hypotheses that enable the automatic recovery of a row basis instead of previous human-guided contributions.

Evaluations across embedded and server-class architectures confirm our method's effectiveness, successfully reconstructing known mappings and uncovering previously unknown scrambling functions.
Our method provides a 99\% recall and accuracy on all tested platforms.
Most notably, \knocknock runs in under a few minutes, even on systems with more than 500GB of DRAM, showcasing the scalability of our method.
Our approach provides an automated, principled pathway to accurate DRAM reverse engineering.
\end{abstract}

\section{Introduction}

Modern computing systems rely on \emph{Dynamic Random Access Memory} (DRAM) as a high-throughput, low-latency memory subsystem.
However, manufacturers often obscure the physical-to-DRAM addressing schemes using undocumented linear scrambling functions~\cite{pessl_drama_2016,wi_sudoku_2025}.
Consequently, reverse engineering becomes a prerequisite for conducting precise side-channel and fault-injection analyses~\cite{kwong_rambleed_2020, frigo_trrespass_2020, bechtel_memory-aware_2022}, as well as for developing effective mitigations~\cite{ZebRAM}. 
Without knowledge of the DRAM addressing function, an attacker cannot reliably target specific DRAM rows or banks, limiting the feasibility and repeatability of DRAM-based attacks such as Rowhammer~\cite{kim_flipping_2014} or memory-aware cache attacks~\cite{bechtel_memory-aware_2022}.
Similarly, defenders cannot deploy targeted mitigations, \eg memory fencing, error detection, or access-pattern randomization~\cite{WiPKKKLA23}, without understanding which physical addresses map to vulnerable DRAM structures~\cite{ZebRAM}.
Beyond security, this mapping is also critical in systems research: memory allocation~\cite{PanGM16, BaiHCL25}, or DRAM-aware data placement strategies~\cite{yoon2011row, wang2020figaro} all benefit from accurate knowledge of the underlying physical-to-DRAM mapping.

Logically, researchers have already tried to reverse engineer this physical-to-DRAM mapping.
Early proposed solutions to reverse DRAM addressing functions either rely on invasive physical probing~\cite{pessl_drama_2016, jung_reverse_2016} or exhaustive brute-force approaches~\cite{pessl_drama_2016, wang_dramdig_2020, jattke_zenhammer_2024}.
This exhaustive search phase has exponential complexity in terms of the number of bits of the DRAM address, which means methods relying on brute force scale poorly with larger DRAM.
As proved by previous work~\cite{wang_dramdig_2020}, exhaustive-search solutions~\cite{pessl_drama_2016} can take hours or even longer to execute successfully on current yet standardly-sized DRAM systems, severely limiting their applicability in practical scenarios.
While some works~\cite{helm_reliable_2020, wi_sudoku_2025} tried addressing this complexity issue to reduce computation time by crafting customized solutions tailored to specific platforms, this came at the expense of generalizability and automation.
Another limitation of existing approaches is noise: The first phase of the reverse engineering process relies on a timing side channel, where misclassification can happen, resulting in broken masks.
While previous works~\cite{pessl_drama_2016, wang_dramdig_2020} repeated the measurements to reduce noise impact, a misclassification in the experimental phase will introduce errors in the reversing phase.

To address these limitations, we present \emph{\knocknock}, a principled, efficient, and entirely automated approach to reverse engineer physical-to-DRAM address translation.
Our approach treats the physical-to-DRAM translation as a black box, reversing the entire pipeline without separating the memory controller and DRAM-internal mappings individually.
\knocknock leverages the row-buffer conflict side channel~\cite{pessl_drama_2016} to build sets of addresses mapping to the same bank or row, then introduces a reduction of the search space using nullspace analysis.
We present, to the best of our knowledge, the first \textbf{analytical framework for physical-to-DRAM address translation}, which we leverage to achieve a more efficient, broadly applicable, and precise analysis.
Our solution eliminates the exponential search phase altogether and introduces four \textbf{contributions}:
\begin{description}
\item[Reduced search space through algebra]
Thanks to an algebraic formalization of the problem, we reduce the search space by proving that the bank/channel masks form a $\operatorname{Nullspace}(D)$ basis, where $D$ is a sparse difference matrix.
Computing that null space is at most of complexity $O(n^3)$ in the number of rows, removing the exponential term that limited prior tools.
\item[Provable, noise-aware sample bound.]
We derived a closed-form expression that theoretically guarantees full mask recovery even with mislabeled samples, turning the heuristic "collect-until-it-works" from previous works into a one-line calculation.
\item[Automatic, low-weight row masks]
From this reduced search space, we retrieve the row addressing function with an algorithm using minor hardware-based hypotheses.
\item[DRAM-agnostic approach]
Because the method needs only the existence of row-buffer conflicts, \knocknock does \emph{not} assume prior knowledge of the DRAM hierarchy (module, rank, bank group, \emph{etc.}) to reverse-engineer the bank masks.
Our method only uses empirically verified assumptions about the hardware to reverse row masks.
This approach extends applicability beyond systems with well-known or documented memory configurations, \eg desktops or HPC-class servers, to more constrained environments like undocumented mobile, IoT SoCs, and cloud platforms.
In these cases, components may be soldered, obfuscated in the PCB, or remote with limited access to system information.
Existing tools often fail under such conditions, or require manual guidance to function effectively~\cite{pessl_drama_2016,helm_reliable_2020}.
\end{description}

Validated on 10 platforms, server, consumer, and embedded SoCs, \knocknock turns DRAM address reverse engineering from a labor-intensive, sometimes platform-specific effort into a generic, noise-resistant push-button procedure that completes in minutes with 99\% accuracy.
Most notably, \knocknock recovered the physical-to-DRAM mapping functions of a server-class SoC with 512 GB of RAM in a few minutes.
To support reproducible and maintainable research, we publish our code and data in an open repository\footnote{\url{https://github.com/antpln/Knock-Knock}}.

The remainder of this paper is organized as follows: 
\Cref{sec:background} presents relevant background and \Cref{sec:related-works} surveys the state-of-the-art.
\Cref{sec:introKnockKnock} introduces the reverse-engineering methodology proposed by Knock-Knock, along with the mathematical formalization used in subsequent sections, which are at the core of our contributions.
Then, \Cref{sec:reverse-engineering} shows how \knocknock reverse engineers the parity masks of the addressing function.
In \Cref{sec:finding-row-bits}, we present our method to retrieve the full row bits of the addresses and generalize our approach to more complex addressing functions.
\Cref{sec:evaluation} describes the experimental setup used to validate our methodology and the results obtained on a range of different types of platforms.
We finally discuss the results obtained in \Cref{sec:discussion} and conclude the paper in \Cref{sec:conclusion}.

\section{Background}\label{sec:background}

\subsection{DRAM Organization}\label{subsec:dram-organization}
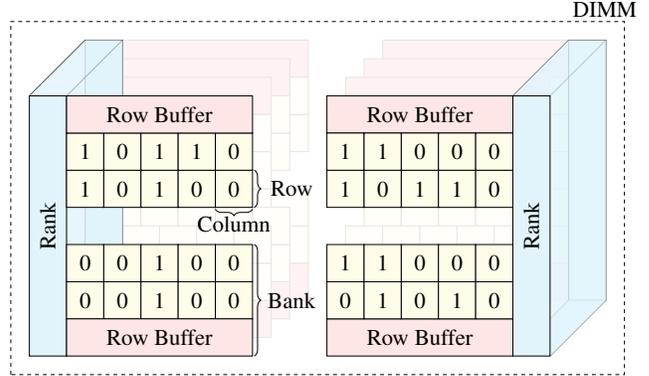
\begin{figure}[t]
\centering
\resizebox{\linewidth}{!}{%
\begin{tikzpicture}

\def\blx{0}
\def\bly{0}

\def\cellh{1}
\def\cellw{1}
\def\rown{2}
\def\coln{5}
\def\rankw{1}

\def\bankn{2}

\def\roww{{\cellw*\coln}}
\def\rankh{7}

\def\depth{0.5}  

\foreach \d in {1.5, 1.0, 0.5} {
  \draw[fill=red!5,draw=black!10] ({\blx + \rankw+\d},{\bly+\d})
    rectangle ({\blx+\rankw+\coln*\cellw+\d}, {\bly+\cellh+\d});
  \draw[fill=red!5,draw=black!10] ({\blx + \rankw+\d},{\bly+\rankh-\cellh+\d})
    rectangle ({\blx+\rankw+\coln*\cellw+\d}, {\bly+\rankh+\d});

  \foreach\row in {1,2,4,5}{
    \foreach\column in {0,1,2,3,4} {
      \draw[fill=yellow!3,draw=black!10]
        ({\blx+\rankw+\column*\cellw+\d}, {\bly+\row*\cellh+\d})
        rectangle ({\blx+\rankw+(\column+1)*\cellw+\d}, {\bly+(\row+1)*\cellh+\d});
    }
  }
}

\draw[fill=cyan!10,draw=black!70] ({\blx+1.5},{\bly+1.5})
  rectangle ({\blx+\rankw+1.5},{\bly + \rankh+1.5});

\draw[black] (\blx,\bly) -- ({\blx+1.5},{\bly+1.5});
\draw[black] ({\blx+\rankw},\bly) -- ({\blx+\rankw+1.5},{\bly+1.5});
\draw[black] (\blx,{\bly+\rankh}) -- ({\blx+1.5},{\bly+\rankh+1.5});
\draw[black] ({\blx+\rankw},{\bly+\rankh}) -- ({\blx+\rankw+1.5},{\bly+\rankh+1.5});
\fill[cyan!10,opacity=0.8] 
    (\blx,{\bly+\rankh}) --
    ({\blx+\rankw},{\bly+\rankh}) --
    ({\blx+\rankw+1.5},{\bly+\rankh+1.5}) --
    ({\blx+1.5},{\bly+\rankh+1.5}) -- cycle;
\fill[cyan!10,opacity=0.8] 
    ({\blx+\rankw},{\bly+\rankh}) --
    ({\blx+\rankw+1.5},{\bly+\rankh+1.5}) --
    ({\blx+\rankw+1.5},{\bly+1.5}) --
    ({\blx+\rankw},{\bly+1.5}) -- cycle;


\draw[fill=cyan!10,draw=black] (\blx,\bly)
  rectangle ({\blx+\rankw},{\bly + \rankh}) node[midway, rotate=90] {\LARGE Rank};

\foreach\row in {1,2,4,5}{
  \foreach\column in {0,1,2,3,4} {
    \pgfmathrandominteger{\randbit}{0}{1}
    \draw[fill=yellow!10,draw=black]
      ({\blx+\rankw+\column*\cellw}, {\bly+\row*\cellh})
      rectangle ({\blx+\rankw+(\column+1)*\cellw}, {\bly+(\row+1)*\cellh})
      node[midway] {\LARGE \randbit};
  }
}

\draw[fill=red!10,draw=black] ({\blx + \rankw},\bly)
  rectangle ({\blx+\rankw+\coln*\cellw}, {\bly+\cellh})
  node[midway] {\LARGE Row Buffer};
\draw[fill=red!10,draw=black] ({\blx + \rankw},{\bly+\rankh-\cellh})
  rectangle ({\blx+\rankw+\coln*\cellw}, {\bly+\rankh})
  node[midway] {\LARGE Row Buffer};

\def\blx{8}
\def\bly{0}

\foreach \d in {1.5, 1.0, 0.5} {
  \draw[fill=red!5,draw=black!10] ({\blx+\d},{\bly+\d})
    rectangle ({\blx+\coln*\cellw+\d}, {\bly+\cellh+\d});
  \draw[fill=red!5,draw=black!10] ({\blx+\d},{\bly+\rankh-\cellh+\d})
    rectangle ({\blx+\coln*\cellw+\d}, {\bly+\rankh+\d});

  \foreach\row in {1,2,4,5}{
    \foreach\column in {0,1,2,3,4} {
      \draw[fill=yellow!3,draw=black!10]
        ({\blx+\column*\cellw+\d}, {\bly+\row*\cellh+\d})
        rectangle ({\blx+(\column+1)*\cellw+\d}, {\bly+(\row+1)*\cellh+\d});
    }
  }
}

\draw[fill=cyan!5,draw=black!70] ({\blx+(\coln*\cellw)+1.5},{\bly+1.5})
  rectangle ({\blx+(\coln*\cellw)+\rankw+1.5},{\bly + \rankh+1.5});

\draw[black!70] ({\blx+(\coln*\cellw)},\bly) -- ({\blx+(\coln*\cellw)+1.5},{\bly+1.5});
\draw[black!70] ({\blx+(\coln*\cellw)},\bly+\rankh) -- ({\blx+(\coln*\cellw)+1.5},{\bly+\rankh+1.5});
\draw[black!70] ({\blx+(\coln*\cellw)+\rankw},\bly) -- ({\blx+(\coln*\cellw)+\rankw+1.5},{\bly+1.5});
\draw[black!70] ({\blx+(\coln*\cellw)+\rankw},{\bly+\rankh}) -- ({\blx+(\coln*\cellw)+\rankw+1.5},{\bly+\rankh+1.5});

\fill[cyan!10,opacity=0.8] 
    ({\blx+(\coln*\cellw)},{\bly+\rankh}) --
    ({\blx+(\coln*\cellw)+\rankw},{\bly+\rankh}) --
    ({\blx+(\coln*\cellw)+\rankw+1.5},{\bly+\rankh+1.5}) --
    ({\blx+(\coln*\cellw)+1.5},{\bly+\rankh+1.5}) -- cycle;
\fill[cyan!10,opacity=0.8] 
    ({\blx+(\coln*\cellw)},{\bly}) --
    ({\blx+(\coln*\cellw)+\rankw},{\bly}) --
    ({\blx+(\coln*\cellw)+\rankw+1.5},{\bly+1.5}) --
    ({\blx+(\coln*\cellw)+1.5},{\bly+1.5}) -- cycle;
\fill[cyan!10,opacity=0.8] 
    ({\blx+(\coln*\cellw)+\rankw},{\bly+\rankh}) --
    ({\blx+(\coln*\cellw)+\rankw+1.5},{\bly+\rankh+1.5}) --
    ({\blx+(\coln*\cellw)+\rankw+1.5},{\bly+1.5}) --
    ({\blx+(\coln*\cellw)+\rankw},{\bly}) -- cycle;


\draw[fill=cyan!10,draw=black] ({\blx+(\coln*\cellw)},\bly)
  rectangle ({\blx+(\coln*\cellw)+\rankw},{\bly + \rankh})
  node[midway, rotate=90] {\LARGE Rank};

\draw[fill=red!10,draw=black] ({\blx},\bly)
  rectangle ({\blx+\coln*\cellw}, {\bly+\cellh})
  node[midway] {\LARGE Row Buffer};
\draw[fill=red!10,draw=black] ({\blx},{\bly+\rankh-\cellh})
  rectangle ({\blx+\coln*\cellw}, {\bly+\rankh})
  node[midway] {\LARGE Row Buffer};

\foreach\row in {1,2,4,5}{
  \foreach\column in {0,1,2,3,4} {
    \pgfmathrandominteger{\randbit}{0}{1}
    \draw[fill=yellow!10,draw=black]
      ({\blx+\column*\cellw}, {\bly+\row*\cellh})
      rectangle ({\blx+(\column+1)*\cellw}, {\bly+(\row+1)*\cellh})
      node[midway] {\LARGE \randbit};
  }
}

\draw[dashed] (-0.5,-0.5) rectangle (16,9) node[above, xshift=-15pt] {\LARGE DIMM};

\draw [decorate,decoration={brace,amplitude=5pt,raise=0.3ex}]
(6,4) -- (5,4) node[midway, yshift = -13pt, thick] {\LARGE Column};

\draw [decorate,decoration={brace,amplitude=5pt,raise=0.3ex}]
(6,5) -- (6,4) node[midway, xshift = 30, thick] {\LARGE Row};

\draw [decorate,decoration={brace,amplitude=5pt,raise=0.3ex}]
(6,3) -- (6,0) node[midway, xshift = 30, thick] {\LARGE Bank};

\end{tikzpicture}
}
\caption{Organization of an example DRAM module. It is composed of two ranks, each handling 8 banks. Each bank is composed of 5 columns and 2 rows. Memory systems may contain several channels, each containing multiple modules.}
\label{fig:dram_organization}
\end{figure}

\begin{figure}[t]
  \centering
  \begin{tikzpicture}
    \begin{axis}[
      xlabel = {Elapsed cycles},
      ylabel = {Occurrences},
      ybar,                      
      bar width = 1,             
      xtick distance = 20,            
      enlarge x limits = 0.1,
      enlarge y limits = 0.05,
      grid = both,
    ]
      \addplot table[
        x       = value,
        y       = freq,
        col sep = comma,
      ] {figures/elapsed_hist_bins.csv};
      \draw[red!80,dashed]
      (axis cs:170,0) rectangle (axis cs:180,550000);
        \node[align=center,anchor=west,font=\normalsize]
      at ($(axis cs:180,0)!0.5!(axis cs:180,550000)$)
      {Low‑latency};
        \draw[blue!80,dashed]
      (axis cs:220,0) rectangle (axis cs:240,120000);
        \node[align=center,anchor=west,font=\normalsize]
      at ($(axis cs:240,0)!0.5!(axis cs:240,120000)$)
      {High‑latency\\ \textbf{Conflict}};
    \end{axis}
  \end{tikzpicture}
  \caption{Distribution of elapsed cycles between loads with and without row conflicts}
  \label{fig:histogram}
\end{figure}
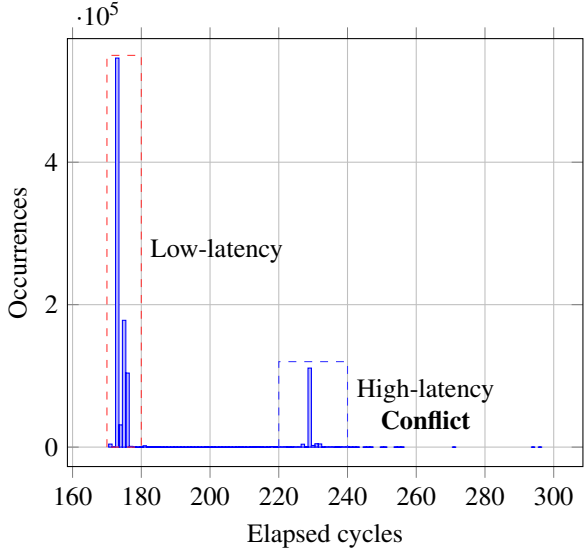

As illustrated in \Cref{fig:dram_organization}, DRAM modules are organized in a hierarchy of channels, modules, ranks, banks, rows, and columns, down to individual cells, each containing a single bit~\cite{ddr4standard, lpddr4standard}.
Each channel can be used independently, allowing access distribution.
These channels can contain multiple modules, which can be further divided into ranks, often one on the front and one on the back of the module.
Those ranks are composed of different row/column arrays, called banks.
Besides the memory array, each bank contains circuits to serve the individual cells, notably the \emph{row buffer}.
From a logical perspective, the row buffer can be seen as a cache for the most recently accessed row.
Subsequent accesses to the same row are directly served from this buffer, reducing access latency.
However, if a \emph{different row is accessed in the same bank}, the DRAM controller must first close the current row and open the new one, which incurs a higher latency.
\Cref{fig:histogram} shows that the access latency follows a mixture of two distributions: one with lower latency when accessing the same row (on the left), and one with higher latency when accessing a different row (on the right).
This event is called a \emph{row conflict} \label{buffer-conflict} and constitutes a fundamental side channel to enable improved Rowhammer~\cite{pessl_drama_2016, kwong_rambleed_2020, frigo_trrespass_2020, kaur_flipping_2023, kogler_half-double_2022} and cache~\cite{bechtel_memory-aware_2022,schwarz_malware_2019} attacks.

\subsection{DRAM Addressing}

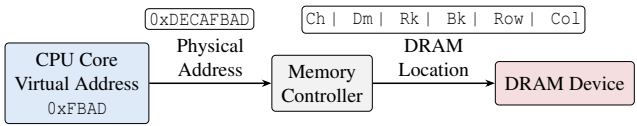
\begin{figure}[t]
\centering
\resizebox{\linewidth}{!}{
\definecolor{coreBlue}{RGB}{ 70,140,210}%
\definecolor{extGray} {RGB}{180,180,180}%
\definecolor{dramRed} {RGB}{200, 70,  80}%
\definecolor{colCh}{RGB}{0,0,0}%
\definecolor{colDm}{RGB}{0,0,0}%
\definecolor{colRk}{RGB}{0,0,0}%
\definecolor{colBk}{RGB}{0,0,0}%
\definecolor{colRow}{RGB}{0,0,0}%
\definecolor{colCol}{RGB}{0,0,0}%
\tikzset{
  bigbox/.style  ={draw,rounded corners,thick,align=center,
                   minimum width=3.3cm,minimum height=1.6cm,
                   fill=#1!18, font=\huge, inner sep=10pt},
  smallbox/.style={draw,rounded corners,thick,align=center,
                   font=\huge,fill=#1!10,inner sep=4pt},
  arrow/.style   ={ultra thick,-Stealth},
  lab/.style     ={font=\huge},
}%
\begin{tikzpicture}[node distance=4.5cm,lab,trim left=(core.west)]

\node[bigbox=coreBlue] (core) {CPU Core\\ Virtual Address\\ \ttfamily 0xFBAD};

\node[bigbox=extGray,right=of core] (mcu) {Memory\\Controller};

\node[bigbox=dramRed,right=of mcu,minimum width=4.3cm] (dram)
      {DRAM Device};

\draw[arrow] (core.east) -- node[midway,above, yshift=3pt,align=center]{Physical \\ Address} (mcu.west);
\draw[arrow] (mcu.east) -- node[midway,above,yshift=3pt,align=center]{DRAM \\ Location} (dram.west);

\node[smallbox=white,above=9mm of $(mcu.north)!0.5!(dram.north)$] (fields)
      {\centering \ttfamily \textcolor{colCh}{Ch}\,|\,
       \textcolor{colDm}{Dm}\,|\,
       \textcolor{colRk}{Rk}\,|\,
       \textcolor{colBk}{Bk}\,|\,
       \textcolor{colRow}{Row}\,|\,
       \textcolor{colCol}{Col}};
       
\node[smallbox=white,above=7mm of $(core.north)!0.5!(mcu.north)$, align=center] (phyaddr)
      {\centering \ttfamily 0xDECAFBAD};

\end{tikzpicture}
}
\caption{Simplified view of translations from Virtual Address to DRAM Location containing the channel, module, rank, bank, row, and column of the accessed byte}
\label{fig:translations}
\end{figure}

\begin{figure}[t]
\centering
\resizebox{0.9\linewidth}{!}{%
\begin{tikzpicture}[%
    bit/.style   ={draw,minimum width=8mm,minimum height=8mm,
                   font=\normalsize,align=center},
    pa/.style    ={bit,fill=blue!10},
    ri/.style    ={bit,fill=orange!20},
    xor/.style   ={draw,circle,inner sep=0pt,minimum size=5mm},
    >=Stealth,   
]

\newcommand{\halfConnector}[2]{%
  \draw[->]
    let
      \p1 = (#1),       
      \p2 = (#2),       
      \n1 = {(\y1+\y2)/2}   
    in
      (\p1)                       
      -- (\x1,\n1)                
      -- (\x2,\n1)                
      -- (\p2);                   
}

\foreach \i/\val in {0/1,1/0,2/1}
  \node[ri] (r\i) at (\i*3.15,2.5) {\val\\ $b_{\i}$};

\node[anchor=north] (bi) at ($(r1.north)+(0,.7)$)
    {\large Bank Index};

\foreach \i/\val in {0/0,1/1,2/0,3/1,4/1,5/1}
  \node[pa] (p\i) at (\i*1.26,0) {\val\\ $p_{\i}$};

\node[anchor=south] at ($(bi.south)+(0,-4.4)$)
    {\large Physical Address};

\node[xor] (x0) at ($(r0)!0.5!(p0 -| r0)$) {};
\halfConnector{p0.north}{x0.south}
\halfConnector{p1.north}{x0.south}
\draw[->] (x0.north) -- (r0.south);

\node[xor] (x1) at ($(r1)!0.5!(p2 -| r1)$) {};
\halfConnector{p2.north}{x1.south}
\halfConnector{p3.north}{x1.south}
\halfConnector{p4.north}{x1.south}
\draw[->] (x1.north) -- (r1.south);

\node[xor] (x2) at ($(r2)!0.5!(p5 -| r2)$) {};
\halfConnector{p5.north}{x2.south}
\draw[->] (x2.north) -- (r2.south);

\node at (x0) {$\oplus$};
\node at (x1) {$\oplus$};
\node at (x2) {$\oplus$};

\node[anchor=west] at ($(r2.east)+(0.1,0)$)
    {\textbf{= 5}};
\node[anchor=west] at ($(p5.east)+(0.1,0)$)
    {\textbf{= 58}};

\end{tikzpicture}
}
\caption{Example of a linear function to address banks. It uses 3 parity masks. $b_{i}$ and $p_{i}$ represent, respectively, the bank and physical address bits.}
\label{fig:linear_example}
\end{figure}
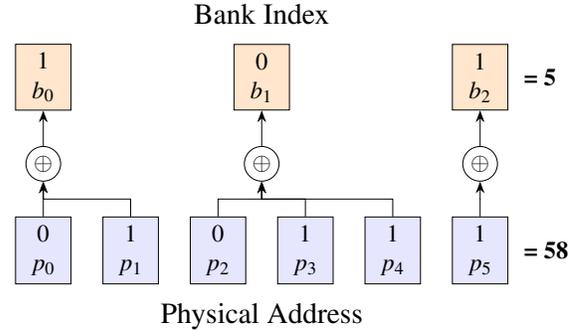

The \emph{memory controller} is responsible for translating \emph{physical addresses} used by the CPU to coordinate system mapping to the location of the data in the DRAM hierarchy. \Cref{fig:translations} summarizes the translations made from virtual addresses to DRAM.
This translation is opaque to any process running on the CPU.
The result of the physical-to-DRAM-location address translation is composed of a 6-tuple \texttt{[Ch, Dm, Rk, Bk, Row, Col]}, which indicates, respectively, the channel, module, rank, bank, row, and column of the target cell.
Previous works showed that this translation is done \emph{linearly}~\cite{kim_flipping_2014,pessl_drama_2016,helm_reliable_2020} by XORing certain bits of the physical address between them to form the DRAM address. An example of such mapping is shown in \Cref{fig:linear_example}, where physical address bits $Pi$ are mapped to bank index bits $Bi$ through XORs.
Which bits are used as inputs for each XOR operator are defined by the so-called \emph{parity masks}.
In some cases~\cite{marazzi2024risc}, the addressing can directly map bits without using XOR functions. This represents a sub-case of linearity, as a direct bit-to-bit assignment is just a parity mask with only one bit set.

This XOR scrambling mechanism allows for optimizations such as bank and channel interleaving \cite{helm_reliable_2020}, reducing both stress on the DRAM cells and access latency.
While AMD publicly documented the functions for older generations of processors in their \emph{BIOS and Kernel Developer's Guide}\cite{pessl_drama_2016}, manufacturers often do not document them publicly.

Therefore, the relation between a physical address and the location of its related byte in the DRAM --- described as the tuple \texttt{[Ch, Dm, Rk, Bk, Row, Col]} --- is often unknown.
\highlightbox{In this work, we address the problem of unknown physical-to-DRAM address mappings and provide a new, provably efficient method to reverse-engineer DRAM address mapping functions that is platform- and DRAM-geometry-agnostic.}

\section{Related Works}
\label{sec:related-works}
In this section, we survey previous related work. 
To better motivate the need to know the physical-to-DRAM mapping, we first (\Cref{sec:attacks-known-mappings}) present different attacks that were only possible after unlocking these addressing functions.
Then (\Cref{sec:rev-eng-mapping-soa}) we introduce important works from the literature, describe their strategies to reverse-engineer address mappings, and provide a qualitative comparison among them and our Knock-Knock proposal.
We finish the review of related works (\Cref{sec:rev-eng-mapping-algebra}) with a deep focus on works that use a linear algebra approach to discover unknown mapping functions.

\subsection{Attacks Through Known Addressing Functions}
\label{sec:attacks-known-mappings}
Various works have already shown how to leverage knowledge of the physical-to-DRAM mapping to craft more powerful attacks.
We can distinguish two main categories:

\textbf{Cache attacks}
While generally not related to DRAM addressing, cache attacks can be enabled or improved by the knowledge of DRAM addressing.
In some cache-based side channels, \eg \FlushReload, the row hits can be confused for cache hits, potentially adding noise for the attack~\cite{pessl_drama_2016}. Thus, taking into account physical-to-DRAM addressing reduces the risk of this confusion.
Bechtel~\etal~\cite{bechtel_memory-aware_2022} built a \emph{DRAM-aware }Denial-of-Service attack on the shared cache of two multi-core embedded platforms.
Knowing the translation from physical addresses, used for cache addressing, to DRAM addressing, the authors force the system's cache misses to target the same DRAM bank.
This greatly increases load latency, as cache misses will also cause a row-buffer conflict and prevent bank-level parallelism.
By using this method, they achieve a slowdown of two orders of magnitude compared to state-of-the-art Denial-of-Service attacks.

Schwarz~\etal~\cite{schwarz_malware_2019} demonstrated an attack targeting RSA implementations executed within Intel's SGX enclaves by constructing cache eviction sets that leverage DRAM address mapping.
The authors start by identifying clusters of addresses in a contiguous memory region (\eg a bank) by using the row-buffer conflict side channel.
Then, the authors use the reversed physical-to-DRAM mapping to recover bits of information about the physical addresses of their set.
Using these recovered eviction sets, they build an eviction set able to attack RSA implementations running inside Intel's SGX security enclave.

\textbf{Rowhammer attacks}
Rowhammer is a micro-architectural fault attack allowing attackers to flip bits outside of the memory allocated to their malicious unprivileged program.
It is an attack on \emph{process memory isolation}.
It takes advantage of a previously known reliability issue of DRAM chips: 
Repeated high-frequency accesses to the same position can leak charge to neighboring cells, possibly causing their values to flip~\cite{kim_flipping_2014}.
Manufacturers have tried to mitigate such attacks, most notably with \emph{Target Row Refresh}.
It uses counters of accesses to DRAM rows in order to refresh victim rows in case of a suspicious number of accesses. 
However, subsequent works have leveraged a known DRAM mapping to find contiguous memory and hence build more complex \emph{hammering patterns} that circumvent TRR either by overloading the counter~\cite{frigo_trrespass_2020} or by using the TRR-induced refreshes to do the hammering~\cite{kogler_half-double_2022}.
Moreover, \emph{RAMBleed}~\cite{kwong_rambleed_2020} showed that precise hammering, requiring in-depth knowledge of the physical-to-DRAM mapping, and the same knowledge of contiguous memory could be used as a read side channel, thus leaking secrets.
This side channel has been developed to broaden its threat model to attacks against DNNs~\cite{RakinCYF22} or cryptographic implementations~\cite{Derya2024}

\subsection{Empirical DRAM Mapping Reverse-Engineering Strategies}
\label{sec:rev-eng-mapping-soa}

\begin{table*}[t]
    \centering
    \caption{Qualitative comparison of \knocknock with representative software-only approaches for DRAM address-mapping recovery.}
    \resizebox{\textwidth}{!}{%
        \rowcolors{2}{rowgray}{white}
\begin{tabular}{@{}lL{3.6cm}L{3.6cm}L{2cm}L{2cm}L{2cm}L{2cm}@{}}
\toprule
\textbf{Tool} & 
\textbf{Core idea} & 
\textbf{Worst-case search complexity} &
\textbf{Provable sample bound} &
\textbf{Noise tolerance} &
\textbf {Platform Agnostic} &
\textbf{Support for complex row addressing}
\\
\midrule

\textbf{DRAMA~\cite{pessl_drama_2016}} &
\begin{minipage}[t]{\linewidth}
\vspace{-\baselineskip}
\begin{titemize}
  \item Cluster conflicts.
  \item Enumerate masks.
\end{titemize}
\end{minipage}
& Exponential in candidate address bits
& \cmarkempty
& \cmarkempty
& \cmarkhalf
& \cmarkempty
\\

\textbf{DRAMDig~\cite{wang_dramdig_2020}} &
\begin{minipage}[t]{\linewidth}
\vspace{-\baselineskip}
\begin{titemize}
  \item Detect row/column bits.
  \item Guided search.
\end{titemize}
\end{minipage}
& Exponential in reduced candidate address bits
& \cmarkempty
& \cmarkempty
& \cmarkempty
& \cmarkempty
\\

\textbf{Helm \etal~\cite{helm_reliable_2020}} &
\begin{minipage}[t]{\linewidth}
\vspace{-\baselineskip}
\begin{titemize}
  \item Cluster conflicts.
  \item Uses Intel counters.
\end{titemize}
\end{minipage}
& Exponential in reduced candidate address bits
& \cmarkempty
& \cmarkfull
& \cmarkempty
& \cmarkempty
\\

\textbf{Sudoku~\cite{wi_sudoku_2025}} &
\begin{minipage}[t]{\linewidth}
\vspace{-\baselineskip}
\begin{titemize}
  \item Cluster conflicts.
  \item 2 timing channels.
\end{titemize}
\end{minipage}
& Exponential in reduced candidate address bits
& \cmarkempty
& \cmarkempty
& \cmarkempty
& \cmarkempty
\\

\textbf{AMDRE~\cite{heckel2023reverse}} &
\begin{minipage}[t]{\linewidth}
\vspace{-\baselineskip}
\begin{titemize}
  \item Cluster conflicts.
  \item Enumerate masks.
\end{titemize}
\end{minipage}
& Exponential in candidate address bits
& \cmarkempty
& \cmarkempty
& \cmarkempty
& \cmarkempty
\\

\textbf{DARE~\cite{jattke_zenhammer_2024}} &
\begin{minipage}[t]{\linewidth}
\vspace{-\baselineskip}
\begin{titemize}
  \item Cluster conflicts.
  \item Enumerate masks.
\end{titemize}
\end{minipage}
& Exponential in candidate address bits
& \cmarkempty
& \cmarkhalf
& \cmarkempty
& \cmarkempty
\\

\textbf{RISC-H~\cite{marazzi2024risc}} &
\begin{minipage}[t]{\linewidth}
\vspace{-\baselineskip}
\begin{titemize}
  \item Cluster conflicts.
  \item Enumerate masks.
\end{titemize}
\end{minipage}
& Exponential in candidate address bits
& \cmarkempty
& \cmarkempty
& \cmarkempty
& \cmarkempty
\\

\textbf{\knocknock~(ours)} & 
\begin{minipage}[t]{\linewidth}
\vspace{-\baselineskip}
\begin{titemize}
  \item Detect conflicts.
  \item Null space analysis.
\end{titemize}
\end{minipage}
& Polynomial through gaussian elimination on $\smash{N\times n}$ matrix
& \cmarkfull
& \cmarkfull
& \cmarkfull
& \cmarkhalf
\\
\bottomrule
\end{tabular}

    }
    \label{tab:method-comparison}
\end{table*}

DRAMA \cite{pessl_drama_2016} introduced the first generic methodology to recover the addressing through the row-buffer-based side channel described in \Cref{buffer-conflict}.
In particular, by identifying which address pairs create a row-buffer conflict, the authors can create a cluster of addresses belonging to the same bank.
As a result, it is now possible to target the same bank, which becomes useful for different attack strategies.
For instance, in the rowhammer case, this enables hammering several lines in the same bank, circumventing existing countermeasures like TRR~\cite{frigo_trrespass_2020}.
In the case of a Denial-of-Service attack on a cache, this allows a potential attacker to force each memory access to be a row conflict, thus increasing latency on top of cache misses~\cite{bechtel_memory-aware_2022}.
DRAMA requires sampling the latencies between pairs of addresses to build as many conflicting sets as the expected total number of banks to find.
Once those sets are built, XOR masks are exhaustively tried until one is found that explains the whole set.
This brute-force approach exhibits exponential complexity with respect to the number of DRAM address bits, resulting in poor scalability on systems with larger memory capacities.
For example, DRAMA requires several hours to compute address masks on systems equipped with 16 GB of DRAM~\cite{wang_dramdig_2020}.

Building upon DRAMA, DRAMDig \cite{wang_dramdig_2020} uses knowledge about the geometry of specific DRAM chips and the processor micro-architecture to reduce the search space of the exponential-time search.
While this made DRAMDig methodology results more \emph{precise} than DRAMA, it severely reduces its portability by making it platform-specific.

Helm~\etal~\cite{helm_reliable_2020} improved DRAMA's methodology by using Intel CPUs' performance counters; specifically, the authors used counters tracking the number of accesses to each channel, rank, and bank.
Similarly to DRAMDig, its gains in performance and reliability come at the cost of generalizability because those counters are only available on a few Intel CPUs, again negatively affecting portability. 

Heckel and Adamsky~\cite{heckel2023reverse} introduced AMDRE, a novel framework tackling DRAM addressing reverse-engineering on AMD platforms that effectively adapts the methodology originally developed in DRAMA.
Jatke~\etal~\cite{jattke_zenhammer_2024} proposed DARE, an AMD-specific technique that exploits enhanced timing-based synchronization and platform-dependent insights.
Notably, they observed that the DRAM address mapping on these systems exhibits non-linear behavior due to physical address remapping.
This means, as a result, that the addressing functions found by previous methods only work on small DRAM clusters but fail on larger, non-consecutive areas.
By introducing a constant offset prior to applying XOR operations, they demonstrated that the mapping can, in some cases, be effectively reduced to a linear form.
Nonetheless, DARE fundamentally relies on an exhaustive brute-force exploration of the XOR mask space.
Marazzi and Razavi~\cite{marazzi2024risc} introduced the first Rowhammer attack on a RISC-V architecture, basing their approach on AMDRE~\cite{heckel2023reverse}.

The Sudoku tool \cite{wi_sudoku_2025} revisits DRAMA's clustering approach and augments it with two additional timing channels using refresh latencies and amplified consecutive-access latencies to \emph{label} each discovered mask with its hierarchy level: channel, rank, bank-group, bank.
While this labeling improves human interpretability, Sudoku does not improve DRAMA and still inherits its exponentially complex mask-enumeration step, requires timing parameters from the memory controller registers, and leaves XOR-scrambled rows unresolved.

\Cref{tab:method-comparison} compiles the existing works that approach the problem of non-algebraic reverse-engineering DRAM address mapping, and provides a comparison among them.
Based on the provided survey of the state of the art, we can see how the field has tackled the problem to improve the performance from the seminal work in DRAMA~\cite{pessl_drama_2016}, by crafting solutions highly adapted to specific platforms. 
These approaches have considered specific Intel processor performance counters~\cite{helm_reliable_2020}, precise knowledge about DRAM geometry~\cite{wang_dramdig_2020}, and adaptations to other platforms like AMD~\cite{heckel2023reverse} or RISC-V~\cite{marazzi2024risc}, as well as improved methodologies~\cite {jattke_zenhammer_2024} based on DRAMA/AMDRE but only for AMD platforms.
As a result:
\highlightbox{We observe how (1) portability has been greatly sacrificed at the cost of performance, and (2) no work has challenged the fundamental problem of reducing the complexity of the brute-force-search approach.}

\subsection{Linear Algebra For Unknown Mapping Functions Discovery}
\label{sec:rev-eng-mapping-algebra}
An algebraic approach to discovering an unknown mapping function consists of formulating the identification of the function as a mathematical problem, and collecting pairs of conflicting addresses to identify such a function while respecting the problem's constraints.
The use of linear algebra finds use also in the construction of mapping functions~\cite{DBLP:journals/tc/VandierendonckB05}. 
Such employment may provide improvements or new reverse-engineering strategies; yet, its scope is different from ours and, for such, we will not discuss the related body of work.

Hofmann \etal~\cite{DBLP:conf/ccs/HofmannFKV24}, from a minimal set of conflicting addresses, recover linear indexing functions through the iterative computation of the basis of the function's nullspace.
The iterative step verifies whether any of the addresses in the conflicting set is also part of the function's kernel.
Simulation-driven analyses, tested on both cache indexing and DRAM bank-indexing functions, show that the approach finds the indexing function with fewer conflict checks with respect to a brute-force approach.

Gerlach \etal~\cite{DBLP:conf/sp/GerlachSFS24} propose an automated and generic approach to reconstruct linear and non-linear mapping functions.
Their approach splits the mapping function $f$ as the composition of several mapping functions $f_{i}$, where $0 \leq i < n$, and $n$ is the bitwidth of the mapping function's output.
Their approach recovers each function $f_{i}$ independently, converting %
each function into a logic formula in \emph{Disjunction Normal Form} (DNF), transforming it to a system of polynomial equations, and retrieving a compact form of this system through the computation of a Gröbner basis for the original system.
This process is then repeated for the subterms of the new system of equations to further reduce the complexity of the final mapping function formula.
Finally, they build the unknown mapping function from the minimized system of equations.

\textbf{Compared to these previous works}, although \knocknock resembles the work of Hofmann \etal~\cite{DBLP:conf/ccs/HofmannFKV24}, it substantially differs in both the approach, the focus, and the results.
Our approach targets the physical-to-DRAM mapping function instead of generic (linear) mapping functions, allowing for the recovery of bank and row mappings.
We base our identification strategy on the computation of the conflicting set's nullspace; as such, we avoid the iterative check and computation of the function's nullspace. 
An extensive validation campaign, encompassing several architectures, supports our methodology, compared to their simulation-only approach that neglects the impact of noisy measurements.
Regarding the work from Gerlach \etal~\cite{DBLP:conf/sp/GerlachSFS24}, the authors recover non-linear functions, whereas we target linear ones.
Nonetheless, Gerlach \etal's generality implies a higher execution time to recover linear functions.
Also, Gerlach \etal need to probe each input bit (\eg the address bits) to know which has an impact on the mapping function's output.
Our algebraic approach inherently identifies such bits (i.e., the mask bits)
Furthermore, the Gröbner base method is quite time-consuming, requiring more than 10 hours to reverse a DRAM function.

In summary, state-of-the-art algebraic approaches lack empirical evaluation, DRAM-specific metrics, and row-mapping reversing.
Consequently, a fully generic, black-box (\ie platform- and DRAM-geometry-agnostic) and faster methodology is needed for more scalable, precise, and efficient discovery of physical-to-DRAM mappings.
This is the open problem that we address with \knocknock{}, where we target an analytical, complexity-bounded, and provable reverse-engineering methodology.
Specifically, the research question (RQ) that we try to answer with this work is:

\begin{tcolorbox}[colback=gray!10]
    \textbf{RQ: Can we generalize the observations from the row-buffer conflict side-channel and derive an analytical approach to design a reverse-engineering methodology portable to the maximum number of targets?}
\end{tcolorbox}%


\section{High-Level Overview of Knock-Knock}
\label{sec:introKnockKnock}

\begin{figure*}[!t]
  \resizebox{\textwidth}{!}{%
      \input{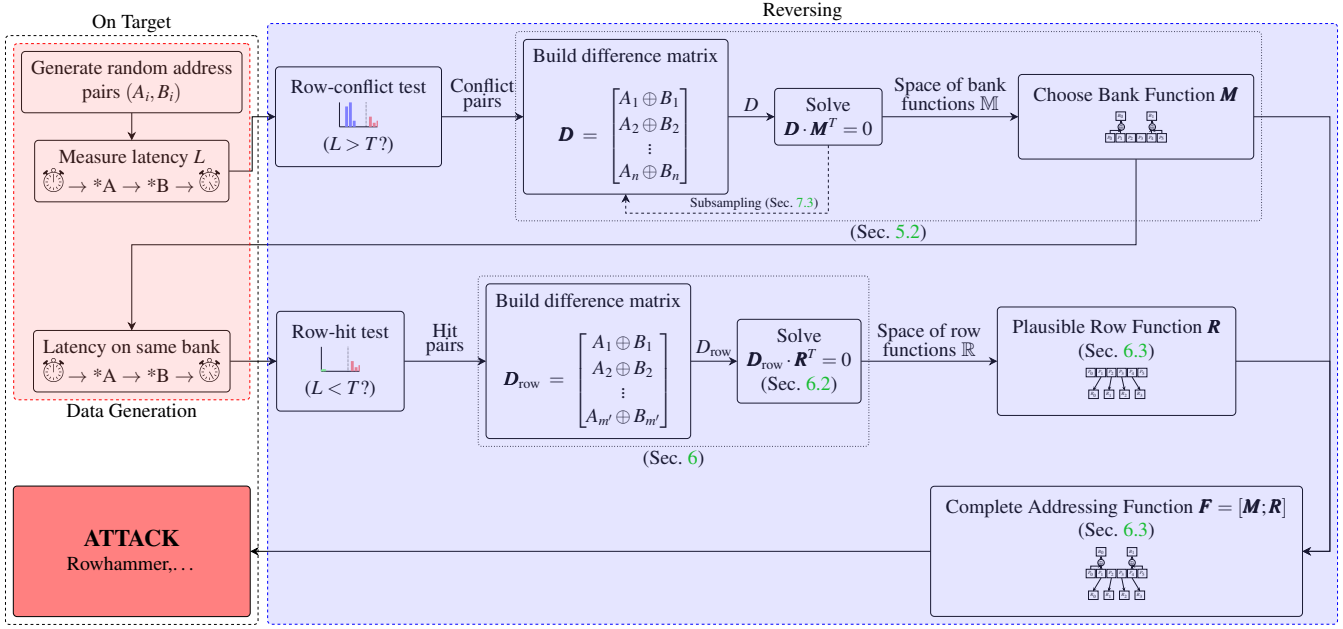}
    }
  \caption{Pipeline of the reverse‑engineering methodology}
  \label{fig:pipeline}
\end{figure*}

In the rest of this paper, we describe \knocknock, our solution for fast, black-box, and platform-agnostic physical-to-DRAM address-mapping reverse engineering.
\Cref{fig:pipeline} illustrates \knocknock{}'s reverse-engineering pipeline, which is divided into two phases: \emph{data generation} (executed on the target system) and \emph{reversing} (performed offline). 
Within the figure, for the reader's convenience, we add references to sections where important parts of the methodology are described.

In the first data generation phase, the system generates random address pairs \((A, B)\) and measures the access latency between them.
Pairs with a high-latency access are clustered as conflicting, \ie belonging to different rows of the same bank, whereas low-latency pairs are labeled as non-conflicting, \ie belonging to different banks or the same row of the same bank.
This phase is identical to the clustering phase of DRAMA~\cite{pessl_drama_2016}.

\knocknock then enters its first reversing phase, described in \Cref{sec:reverse-engineering}.
Then, we use the conflicting pairs to build a difference matrix, enabling the recovery of the \emph{bank addressing function} by solving a system of linear equations.

The bank mappings are then used in the second data generation phase to leverage a variant of the row-buffer side channel to build pairs of addresses belonging to different rows in the same bank.
The system generates address pairs \((A, B)\) \emph{belonging to the same bank} and measures the access latency between them.
Pairs with a high-latency access denote a row-buffer conflict, \ie belonging to different rows.
On the contrary, pairs with a low access latency indicate a row-buffer hit, \ie both addresses belong to the same row of the same bank.

The second reversing phase, described in \Cref{sec:finding-row-bits}, uses the cluster of addresses belonging to the same row of the same bank to construct another difference matrix to derive the \emph{row addressing function}.
The inferred bank and row functions form the \emph{complete physical-to-DRAM addressing function}.

\subsection{Using Linear Algebra for Reverse-Engineering}

The fundamental idea behind \knocknock is to formulate the problem of identifying the parity masks as a linear algebra problem.
We first mathematically define when two addresses cause a row-buffer conflict. Second, from this definition and from a set of randomly generated address pairs, we show how to build a system of linear equations that puts conflicting addresses in relation. Third, we prove that the solution to this system is a valid set of parity masks.
By describing the search problem as the computation of a solution to a system of linear equations, we effectively bound the time complexity of the worst-case scenario to the time complexity of the particular algorithm used to solve the linear system.
While this reduction is sufficient for the channel and bank masks, for which we only want to check equality between addresses, we later introduce some different hypotheses to determine functions for the rows.

An important assumption supporting our methodology is the \emph{linearity} of the address mapping function, which many previous works verified in practice in different instances~\cite{pessl_drama_2016,wang_dramdig_2020,helm_reliable_2020,kogler_half-double_2022}.
To the best of our knowledge, only ZenHammer \cite{jattke_zenhammer_2024} showed a case of non-linear mapping, implied by an offset added to physical addresses.
We believe that our methodology is fully applicable even in such cases: As ZenHammer's authors showed, it is possible to remove this offset, making the mapping linear.

\subsection{Mathematical Definitions and Notation}\label{sec:formalization}
We denote scalar variables with small italic letters (e.g., $\scalar{n}$ number of bits).
Capital italic letters denote $\scalar{n}$-bit vectors (e.g., a physical address $\addr{A}$); we also use the notation $\addr{A}_{\scalar{i}}$ to define several vectors, where the scalar index $\scalar{i}$ is bound by the context.
A capital bold italic letter denotes a matrix (e.g., $\mymatrix{M}$) defined on the set $\{0, 1\}$.
We use a blackboard bold capital letter to denote a set (e.g., a set $\myset{S}$).
The operator $\card{\cdot}$ defines the cardinality of a set (e.g., $\card{\myset{S}}$).

We consider a DRAM addressed with $\scalar{n}$-bit addresses, and describe its locations with 6-tuples $\langle \text{Channel},\text{module},\text{Rank}, \text{Bank}, \text{Row}, \text{Column} \rangle$. For a given physical address $\addr{A}$, we use the notation $\operatorname{Row}(\addr{A})$, $\operatorname{Bank}(\addr{A})$, $\operatorname{Channel}(\addr{A})$ to extract the respective components of $\addr{A}$'s DRAM location.
Given $\Phi$ and $\Delta$, the DRAM's address space and location space, respectively, we define
\begin{equation}
  f : \Phi \to \Delta
  \label{eq:mapping-function}
\end{equation}
as the DRAM  address mapping function.

We denote the access latency for two physical addresses $\addr{A}$ and $\addr{B}$ with $L(\addr{A}, \addr{B})$.
As explained in \Cref{subsec:dram-organization}, for two randomly chosen addresses $\addr{A}$ and $\addr{B}$, the access latency follows a mixture of two distributions.
We define $T$ as the threshold latency that separates the two distributions.
We denote with $\land$, according to the context, either the bit-wise or the logical AND. 
Then, we can describe the relation between row-buffer conflict and access latency with the following proposition:

\begin{align*}
L(\addr{A}, \addr{B}) > T \Leftrightarrow
\operatorname{Row}(\addr{A}) \neq \operatorname{Row}(\addr{B}) & \land \operatorname{Bank}(\addr{A}) = \operatorname{Bank}(\addr{B})\\  &\land \operatorname{Chan}(\addr{A}) = \operatorname{Chan}(\addr{B})
\end{align*}

Denoting with $\oplus$ the bit-wise XOR and with $(\addr{A})_{\scalar{l}}$ the access to the $\scalar{l}$-th bit of the $\scalar{n}$-bit address $\addr{A}$, we define

\begin{equation}
p(\addr{A}) = \bigoplus_{\scalar{l} = 0}^{\scalar{n} - 1}(\addr{A})_{\scalar{l}}
\label{eq:parity}
\end{equation}

the parity of $\scalar{A}$.

For any matrix $\mymatrix{M}$, we define with $\operatorname{rk}(\mymatrix{M})$ the number of \emph{linearly independent} rows (equivalently, columns) of $\mymatrix{M}$.
With $\mymatrix{M}^{T}$ we denote the \emph{transpose} of $\mymatrix{M}$.

We use $\operatorname{HW}(\addr{V})$ as the Hamming weight of the bit vector $\addr{V}$.
We denote with $P(\cdot)$ the probability of a certain event.

\section{Finding Bank and Channel Parity Masks via Nullspace Analysis}\label{sec:reverse-engineering}

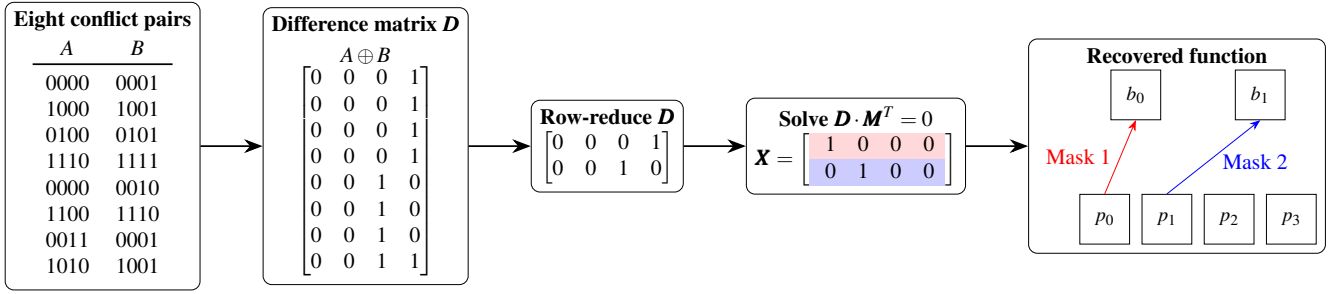
\begin{figure*}[t]
\centering
\resizebox{\linewidth}{!}{
    \begin{tikzpicture}[
  box/.style = {draw, rounded corners, inner sep=4pt, font=\normalsize},
  arr/.style = {-{Stealth[length=3mm]}, thick},
]
\node[box, align=center] (pairs) {%
  \textbf{Eight conflict pairs}\\[2pt]
  \begin{tabular}{cc}
  $\addr{A}$ & $\addr{B}$\\\toprule
  0000 & 0001\\
  1000 & 1001\\
  0100 & 0101\\
  1110 & 1111\\
  0000 & 0010\\
  1100 & 1110\\
  0011 & 0001\\
  1010 & 1001
  \end{tabular}
};
\node[box, align=center, right=1cm of pairs] (D) {%
  \textbf{Difference matrix $\mymatrix{D}$}\\[2pt]
  $\addr{A} \oplus \addr{B}$\\
  \(\displaystyle
    \begin{bmatrix}
    0&0&0&1\\
    0&0&0&1\\
    0&0&0&1\\
    0&0&0&1\\
    0&0&1&0\\
    0&0&1&0\\
    0&0&1&0\\
    0&0&1&1
    \end{bmatrix}\)
};

\node[box, align=center, right=1cm of D] (rref) {%
  \textbf{Row‑reduce $\mymatrix{D}$}\\[2pt]
  \(\displaystyle
    \begin{bmatrix}
    0&0&0&1\\
    0&0&1&0
    \end{bmatrix}\)
};
\node[box, align=center, right=1cm of rref] (null) {%
  \textbf{Solve $\mymatrix{D} \cdot \mymatrix{M}^{T} = 0$}\\[2pt]
    \(\mymatrix{X} = 
  \left[
  \begin{array}{cccc}
    \rowcolor{red!15}\hspace{2pt}1 & 0 & 0 & 0\\
    \rowcolor{blue!20}\hspace{2pt}0 & 1 & 0 & 0
  \end{array}
  \right]
  \)
};

\node[box, align=center, right=1cm of null] (masks) {%
  \textbf{Recovered function}\\[2pt]
    \begin{tikzpicture}[%
    bit/.style   ={draw,minimum width=8mm,minimum height=8mm,
                   font=\small,align=center, sharp corners},
    pa/.style    ={bit},
    ri/.style    ={bit},
    xor/.style   ={draw,circle,inner sep=0pt,minimum size=5mm},
    >=Stealth,   
    ]
        \foreach \i in {0,1,2,3}
            \node[pa] (p\i) at (\i,0) {$\scalar{p}_{\i}$};
        \foreach \i in {0,1}
            \node[ri] (b\i) at (\i*2 + 0.5,2) {$\scalar{b}_{\i}$};
        \draw[->, red] (p0.north) -- node[left]{Mask 1}(b0.south);
        \draw[->, blue] (p1.north) -- node[right, yshift=-2pt]{Mask 2}(b1.south);
    \end{tikzpicture}
};

\draw[arr] (pairs) -- (D);
\draw[arr] (D) -- (rref);
\draw[arr] (rref) -- (null);
\draw[arr] (null) -- (masks);

\end{tikzpicture}
}
\caption{The eight conflict pairs $(\addr{A},\addr{B})$ only differ in their two low-order bits, so every XOR difference $\addr{A} \oplus \addr{B}$ contains at least 0001 or 0010. Stacking those differences produces the matrix $\mymatrix{D}$. We eliminate rows to only keep independent rows. We solve $\mymatrix{D} \cdot \mymatrix{M}^{T} = 0$ and get a basis, giving us the functions that associate the physical address bits $\scalar{p}_{\scalar{i}}$ to the bank index bits $\scalar{b}_{\scalar{i}}$.}
\label{fig:toy_example}
\end{figure*}

This section tackles the first reverse engineering phase of \knocknock, illustrated in \Cref{fig:toy_example}.
The goal is to retrieve bank and channel parity masks in a purely platform-agnostic approach.
While our approach does not separate which bits are used to address the channel and which to address the banks, this mask allows the translation from a physical address to a unique bank.
By using the clusters of pairs of addresses belonging to the same bank, we build a difference matrix, \ie a matrix containing the results of the XOR of each pair of addresses.
This matrix is then used to produce a system of linear equations.
The solution to this system gives a basis for the bank bits of the physical address.

\subsection{Problem statement}

The problem of reverse-engineering the bank and channel parity masks is to find a set of $\scalar{k} \geq 1$ masks $\addr{M}_j \in \{0,1\}^\scalar{n}$ such that, for any two physical addresses $\addr{A}_\scalar{i}, \addr{B}_\scalar{i} \in \{0,1\}^\scalar{n}$ in the same memory module\footnote{In order to simplify notations, we do not include memory module in the equation.}, the following condition holds $\forall \scalar{j} \in [\,1, \scalar{k}\,]$:
\begin{align}
  \begin{split}
    p(\addr{A}_\scalar{i} \land \addr{M}_\scalar{j}) = p(\addr{B}_\scalar{i} \land \addr{M}_\scalar{j})
    \Leftrightarrow
    \operatorname{Bank}(\addr{A}_\scalar{i}) = \operatorname{Bank}(\addr{B}_\scalar{i})\\
    \;\land\;
    \operatorname{Chan}(\addr{A}_\scalar{i}) = \operatorname{Chan}(\addr{B}_\scalar{i})
    \label{eq:rev-eng-conflict-prop}
  \end{split}
\end{align}

That is, the parity distinguishes which addresses are in the same bank and channel.

\subsection{Reduction to Nullspace Analysis}
\label{sec:nullspace-analysis}
Let us consider a set $\myset{S}$ of $\scalar{m} \geq 1$ randomly generated address pairs $\addr{A}_{\scalar{i}}, \addr{B}_{\scalar{i}}$:

\begin{equation}
    \myset{S} = \{(\addr{A}_{\scalar{i}}, \addr{B}_{\scalar{i}}) \in \Phi^2 \mid \operatorname{Bank}(\addr{A}_{\scalar{i}}) = \operatorname{Bank}(\addr{B}_{\scalar{i}}) \land \operatorname{Chan}(\addr{A}_{\scalar{i}}) = \operatorname{Chan}(\addr{B}_{\scalar{i}})\}
    \label{eq:rev-eng-proof-address-set}
\end{equation}

Using the definition of parity (\Cref{eq:parity}), we refactor the \emph{if} term of \Cref{eq:rev-eng-conflict-prop} as:

\[
\bigoplus_{\scalar{l} = 0}^{\scalar{n} - 1} (\addr{A}_{\scalar{i}} \land \addr{M}_{\scalar{j}} \oplus \addr{B}_{\scalar{i}} \land \addr{M}_{\scalar{j}})_{\scalar{l}} = 0,
\]

and since $\land$ is distributive over $\oplus$, we have that:

\begin{equation}
\bigoplus_{\scalar{l} = 0}^{\scalar{n} - 1} ((\addr{A}_{\scalar{i}}\oplus \addr{B}_{\scalar{i}}) \land \addr{M}_{\scalar{j}})_{\scalar{l}} = 0.
\label{eq:rev-eng-proof-refactor-distributive}
\end{equation}

Given $\addr{D}_{\scalar{i}} = \addr{A}_{\scalar{i}} \oplus \addr{B}_{\scalar{i}}$ the $\scalar{i}$-\emph{th} \emph{difference word}, we rewrite \Cref{eq:rev-eng-proof-refactor-distributive} as:
\begin{equation}
  \bigoplus_{\scalar{l} = 0}^{\scalar{n} - 1} (\addr{D}_{\scalar{i}})_{\scalar{l}} \land (\addr{M}_{\scalar{j}})_{\scalar{l}} = \addr{D}_{\scalar{i}} \cdot \addr{M}_{\scalar{j}}^{T} = 0.
  \label{eq:rev-eng-proof-dot-prod}
\end{equation}

Then, we can define the following difference matrix:
\[
\mymatrix{D} =
\begin{pmatrix}
\addr{D}_{1} \\
\addr{D}_{2} \\
\vdots \\
\addr{D}_{\scalar{m}}
\end{pmatrix}
\in \{0,1\}^{\scalar{m} \times \scalar{n}}.
\]

For $\scalar{k}$ parity masks, we define the mask matrix:
\[
\mymatrix{M} =
\begin{pmatrix}
\addr{M}_{1} \\
\addr{M}_{2} \\
\vdots \\
\addr{M}_{\scalar{k}}
\end{pmatrix}
\in \{0,1\}^{\scalar{k} \times \scalar{n}}.
\]

Given $\mymatrix{D}$ and $\mymatrix{M}$, we can rewrite the \emph{if} part of \Cref{eq:rev-eng-conflict-prop} in matrix form:
\[
\mymatrix{D} \cdot \mymatrix{M}^{T} = 0.
\]

Under the condition of sufficient $\scalar{m}$ number of address pairs (see \Cref{subsec:required-samples}), the Rank-Nullity theorem~\cite{katznelson_terse_2008} guarantees that the nullspace of the difference matrix $\mymatrix{D}$ coincides with the set $\myset{M}$ of possible parity masks:

\begin{equation}
  \operatorname{nullspace}(\mymatrix{D}) = \{ \addr{M}_{\scalar{j}} \in \{0,1\}^{\scalar{n}} \mid \mymatrix{D} \cdot \addr{M}_{\scalar{j}}^{T} = 0 \}.
  \label{eq:rev-eng-proof-nullspace-equiv}
\end{equation}

Thus, we have :
\[
\boxed{
    \forall\ \mymatrix{D} \in \{0,1\}^{\scalar{m} \times \scalar{n}},\ \myset{M} = \operatorname{nullspace}(\mymatrix{D}).
}
\]

Any base of $\myset{M}$ describes, in matrix form, a bank addressing function.
We remark that for a given difference matrix $\mymatrix{D}$, the nullspace basis is not unique in general. However, any two bases \(\mymatrix{B}, \mymatrix{B}' \in \{0,1\}^{\scalar{k} \times \scalar{n}}\) of \(\operatorname{nullspace}(\mymatrix{D})\) classify any address pair in the same manner (\ie as conflicting or non-conflicting addresses) under the condition that there exists an invertible matrix \(\mymatrix{P} \in \{0, 1\}^{\scalar{k} \times \scalar{k}}\) such that $\mymatrix{B} = \mymatrix{P} \cdot \mymatrix{B}'$.
Indeed, given two randomly generated addresses $\addr{A}$ and $\addr{B}$, we can expand \Cref{eq:rev-eng-proof-dot-prod} as:

\begin{align*}
\addr{D} \cdot \mymatrix{B}^{T} = (\addr{A} \oplus \addr{B}) \cdot \mymatrix{B}^{T} = (\addr{A} \oplus \addr{B}) \cdot (\mymatrix{P} \cdot \mymatrix{B}')^{T} = (\addr{A} \oplus \addr{B}) \cdot (\mymatrix{B}')^{T} \cdot \mymatrix{P}^{T}.
\end{align*}

Being $\mymatrix{P}$ invertible and $\addr{D} \cdot \mymatrix{B}^{T} = 0$, we have that:

\begin{align*}
(\addr{A} \oplus \addr{B}) \cdot \mymatrix{B}^{T} = (\addr{A} \oplus \addr{B}) \cdot (\mymatrix{B}')^{T}= 0.
\end{align*}

We have shown how the computation of the nullspace of a matrix built from conflicting addresses provides us with a set of parity masks indexing the bank and channel of an address.
We now show how to recover the masks used to index the row of an address.

\summarybox{Using nullspace analysis, we retrieve a set of parity masks that describe the mapping from physical addresses to banks and channels.}

\section{Finding Row Parity Masks via Nullspace Analysis}\label{sec:finding-row-bits}
Now that we have a set of masks \(\addr{M}_{\scalar{j}}\) determining if two addresses belong to the same bank and channel, we need to identify the masks indexing the row bits in the addresses.
To this end, we apply the same nullspace approach described in \Cref{sec:reverse-engineering}, but we consider the subset of only-conflicting addresses:
Addresses belonging to the same bank have a lower access latency when mapped to the same row (\ie row hits) than when mapped to different rows (\ie row conflicts).
This different use of the row-buffer-conflict side channel corresponds to the second Data Generation phase of \Cref{fig:pipeline}.
The objective of this section is to build a difference matrix from this data cluster, which enables us to build a row addressing function with minimal hardware hypotheses, corresponding to the second reversing phase of \Cref{fig:pipeline}.

\subsection{A Sufficient Condition to Reduce the Row Masks Search}

Using our clusters, we begin by searching a vector of bits $\addr{M}_{\text{row}}$ that never vary between two addresses in the same row.
In fact, if a bit varies in the same row address pair, it cannot be used for row calculation.
Let us consider the set $\myset{S}$ of randomly generated addresses $\addr{A}$ and $\addr{B}$ (\Cref{eq:rev-eng-proof-address-set}).

Thanks to \Cref{eq:rev-eng-conflict-prop} and the bank masks \(\addr{M}_{\scalar{j}}\) we have found, we have that:
\[
   \myset{S} = \{(\addr{A}_{\scalar{i}}, \addr{B}_{\scalar{i}}) \in \Phi^2 \mid \forall\ \scalar{j} \in [1, \scalar{k}], p(\addr{A}_{\scalar{i}} \land \addr{M}_{\scalar{j}}) = p(\addr{B}_{\scalar{i}} \land \addr{M}_{\scalar{j}})\}
\]

Being $T$ the threshold separating low-latency access addresses from the high-latency ones (\Cref{sec:formalization}), we define $\myset{S}_{\text{low}}$ the set of address pairs in $\myset{S}$ mapping to the same row (\ie they have a lower latency access):
\[
    \myset{S}_{\text{low}} = \{(\addr{A}_{\scalar{i}}, \addr{B}_{\scalar{i}}) \in \myset{S} \mid L(\addr{A}_{\scalar{i}}, \addr{B}_{\scalar{i}}) < T\} \subseteq \myset{S}.
\]
As stated above, we look for parity masks $\addr{M}_{\text{row,low}}$ selecting bits that are always the same in both $\addr{A}_{\scalar{i}}, \addr{B}_{\scalar{i}} \in \myset{S}_{\text{low}}$:

\begin{equation}
    \forall\, \scalar{l} \in [0, \scalar{n} -1 ]:\, (\addr{M}_{\text{row,low}})_l = \bigwedge_{(\addr{A}_{\scalar{i}}, \addr{B}_{\scalar{i}}) \in \myset{S}_{\text{low}}} [(\addr{A}_{\scalar{i}})_{\scalar{l}} = (\addr{B}_{\scalar{i}})_{\scalar{l}}].
    \label{eq:row-mask-condition}
\end{equation}

From \Cref{eq:row-mask-condition}, we derive the following \textbf{sufficient condition} $\forall\, \addr{A}_{\scalar{i}}, \addr{B}_{\scalar{i}} \in \myset{S}_{\text{low}}$:

\begin{equation}
  \addr{M}_{\text{row,low}} \land \addr{A}_{\scalar{i}} = \addr{M}_{\text{row,low}} \land \addr{B}_{\scalar{i}} \implies \operatorname{Row}(\addr{A}_{\scalar{i}}) = \operatorname{Row}(\addr{B}_{\scalar{i}}).
  \label{eq:row-mask-sufficient-condition}
\end{equation}

This condition restricts the search of row parity masks $\addr{M}_{\text{row, low}}$ to only those that satisfy \Cref{eq:row-mask-condition}.

A simple row mapping (\ie all the row bits set sequentially in the physical address) will give the complete row mapping.

\summarybox{By comparing same-bank addresses belonging to the same row, we build a first coarse-grained row addressing function.}

\subsection{Building a Basis for the Row Parity Masks}
\label{sec:row-basis}
From the set $\myset{S}_{\text{low}}$ of address pairs mapped to the same bank and row, we create the following difference matrix:

\begin{equation*}
\mymatrix{D}_{\text{row}} =
\begin{pmatrix}
\addr{A}_{1} \oplus \addr{B}_{1} \\
\addr{A}_{2} \oplus \addr{B}_{2} \\
\vdots \\
\addr{A}_{\scalar{m}'} \oplus \addr{B}_{\scalar{m}'} \\
\end{pmatrix}
\in \{0,1\}^{\scalar{m}' \times \scalar{n}},
\end{equation*}

where $\addr{A}_{\scalar{i}},\, \addr{B}_{\scalar{i}} \in \myset{S}_{\text{low}}$, and $\scalar{m}' = \card{\myset{S}_{\text{low}}}$.

Given $\scalar{k}' \geq 1$ row masks, we define the row mask matrix:

\begin{equation*}
\mymatrix{R} =
\begin{pmatrix}
\addr{R}_{1}\\
\addr{R}_{2}\\
\vdots \\
\addr{R}_{\scalar{k}'}\\
\end{pmatrix}
\in \{0,1\}^{\scalar{k}' \times \scalar{n}}.
\end{equation*}

We rewrite the \emph{if} condition of \Cref{eq:row-mask-sufficient-condition} as:

\begin{equation*}
    \mymatrix{D}_{\text{row}} \cdot \mymatrix{R}^{T} = 0.
\end{equation*}

Under the condition that $\myset{S}_{\text{low}}$ contains a sufficiently large number $\scalar{m}'$ of address pairs (see \Cref{subsec:required-samples}), the Rank-Nullity theorem~\cite{katznelson_terse_2008} guarantees that:
\[
\boxed{
    \forall\ \mymatrix{D}_{\text{row}} \in \{0,1\}^{\scalar{m}' \times \scalar{n}},\ \myset{R} = \operatorname{nullspace}(\mymatrix{D}_{\text{row}}).
}
\]

We reduce $\myset{R}$ by removing from it the parity masks that selects bits not causing a row conflict when flipped (\ie, not contributing to the row-index computation).
Thus, we remove the masks that select column bits in the address. 

A basis $\mymatrix{R}$ of $\myset{R}$ is equivalent to the $\mymatrix{R}_{\text{true}}$ basis underlying the real row address mapping, up to a change of basis (\Cref{sec:nullspace-analysis}); that is, given any two addresses $\addr{A}$ and $\addr{B}$, the row parity masks generated by $\mymatrix{R}$ provide the same classification for $\addr{A}$ and $\addr{B}$ (\ie they row-conflict or not) as the $\mymatrix{R}_{\text{true}}$'s row parity masks.

The complete address mapping function $f$ (\Cref{eq:mapping-function}) is defined, in matrix form, as:
\begin{equation*}
    \mymatrix{F} =
    \begin{pmatrix}
        \mymatrix{M}\\
        \mymatrix{R}
    \end{pmatrix}.
\end{equation*}

In the next section, we provide an algorithm to determine such a basis $\mymatrix{R}$.

\begin{figure}[t]
    \centering
    \resizebox{\linewidth}{!}{
        \begin{tikzpicture}[%
    bit/.style   ={draw,minimum width=8mm,minimum height=8mm,
                   font=\small,align=center},
    pa/.style    ={bit,fill=blue!10},
    ri/.style    ={bit,fill=orange!20},
    xor/.style   ={draw,circle,inner sep=0pt,minimum size=5mm},
    >=Stealth,   
]
\foreach \i/\val in {0/0,1/1,2/0,3/1,4/1,5/0,6/0}
  \node[pa] (p\i) at (\i*1.05,0) {\val\\ $p_{\i}$};

\node[anchor=north,font=\normalsize] at ($(p3.south)+(0,-.2)$)
    {Physical address};

\foreach \i/\val in {0/0,1/1,2/0,3/1,4/1,5/0,6/0}
  \node[ri] (r\i) at (\i*1.05,3) {\val\\ $r_{\i}$};

\node[anchor=south,font=\normalsize] at ($(r3.north)+(0,+.2)$)
    {Row index};

\node[xor] (x0) at ($(r0)!0.5!(p0)$) {};
\draw[->] (p0.north) |- (x0.south);
\draw[->] (p1.north) |- (x0.east);
\draw[->] (p2.north) |- (x0.east);
\draw[->] (x0.north) -- (r0.south);

\node[xor] (x2) at ($(r5)!0.5!(p5)$) {};
\draw[->] (p4.north) |- (x2);
\draw[->] (p6.north) |- (x2);
\draw[->] (x2) -- (r5.south);

\node at (x0) {$\oplus$};
\node at (x2) {$\oplus$};

\draw[blue!50, dashed, very thick] (-0.65,-.5) rectangle (2.63, 3.5);
\draw[red!50, very thick] (3.68,-.5) rectangle (6.85, 3.5);

\end{tikzpicture}
    }
    \caption{Two row addressing functions. The one highlighted in blue verifies \cref{item:self-ref-condition} while the one in red does not.
    $r_{i}$ and $p_{i}$ represent, respectively, the row and physical address bits.}
    \label{fig:function_constraints}
\end{figure}
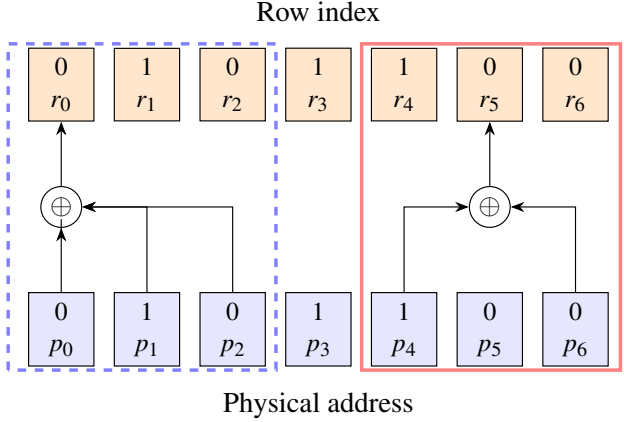

\subsection{An Algorithm to Recover the Row-Mapping Function}
\label{sec:plausible-basis}

We aim to recover a plausible basis \( \mymatrix{R} \in \{0, 1\}^{\scalar{k}' \times \scalar{n}} \) for the row parity verifying the following conditions:

\begin{enumerate}[label=\textbf{C.\arabic*}]
    \item Each row \( \mymatrix{R}_{\scalar{j}} \) includes bit position \( \scalar{j} \), meaning \( \mymatrix{R}_{\scalar{j}}[\scalar{j}] = 1 \), as illustrated in \Cref{fig:function_constraints}
        \label{item:self-ref-condition}. This pivot mirrors the common hardware implementation of a direct wire combined with a few XOR gates and gives us a canonical basis,
    \item The basis has a Hamming weight as low as possible, implying a lower amount of logical gates in hardware. We hypothesize this as a reasonable assumption to minimize resources, which is sustained by previous findings~\cite{pessl_drama_2016, kogler_half-double_2022, jung_reverse_2016},
    
    \item To guarantee that $\mymatrix{F}$ provides an equivalent addressing to the targeted physical-to-DRAM addressing function, the rank of $\mymatrix{R}$ must satisfy:
    \begin{equation*}
        \operatorname{rk}(\mymatrix{F}) = \operatorname{rk}(\mymatrix{M}) +  \operatorname{rk}(\mymatrix{R}).
    \end{equation*}
\end{enumerate}

We discuss the justification and implications of these hypotheses in \Cref{sec:discussion}.
We remark that the linear independence of $\mymatrix{R}$'s rows guarantees the surjectivity of the row addressing function.

\begin{algorithm}[t]
\caption{Rank–aware Row Basis Search}
\label{alg:rank-aware-backtrack}
\begin{algorithmic}[1]
\footnotesize
\Require{
    $\mymatrix{M} \in \{0, 1\}^{\scalar{k} \times \scalar{n}}$ — fixed bank‑mask matrix,\\
    $\myset{R} \subseteq \{0, 1\}^{\scalar{n}}$ — candidate row‑mask vectors,\\
    $\scalar{k}'$ — target row‑space dimension.
}
\Ensure{$\mymatrix{R} \in \{0, 1\}^{\scalar{k}' \times \scalar{n}}$ with
         $\operatorname{rk}[\mymatrix{M};\mymatrix{R}]= \scalar{k} + \scalar{k}'$ and
         minimal total Hamming weight.}
\Statex\Function{Backtrack}{$\scalar{j},\;\mymatrix{B},\;\scalar{r},\;\scalar{w}$}
   \If{$\scalar{r} + (\scalar{k}' - \scalar{j}) < \scalar{r}_{\text{init}} + \scalar{k}'$}        \Comment{rank can no longer reach $\scalar{k} + \scalar{k}'$}\label{line:algo-1-no-full-rank-b}
      \State \textbf{return}\label{line:algo-1-no-full-rank-e}
   \EndIf
   \If{$\scalar{j} = \scalar{k}'$} \Comment{full basis selected}\label{line:algo-1-found-full-basis-b}
      \If{$\scalar{w} < \scalar{w}_{\text{min}}$}
         \State $\mymatrix{B}_{\mathrm{best}}\gets \mymatrix{B}$, \; $\scalar{w}_{\min}\gets \scalar{w}$
      \EndIf
      \State \textbf{return}\label{line:algo-1-found-full-basis-e}
   \EndIf
   
   \ForAll{$\addr{V} \in \myset{C}_{\scalar{j}}$}\label{line:algo-1-search-partion-b}
      \If{$\operatorname{rk}([\mymatrix{M};\mymatrix{B}; \addr{V}]) = \scalar{r}$} \Comment{$\addr{V}$ linearly dependent}\label{line:algo-1-skip-dependent-mask-b}
         \State \textbf{continue}
      \EndIf\label{line:algo-1-skip-dependent-mask-e}
      \State $\mymatrix{B}' \gets [\mymatrix{B}; 
      \addr{V}$],\;
             $\scalar{r}' \gets \scalar{r} + 1$,\;
             $\scalar{w}' \gets \scalar{w} + \operatorname{HW}(\addr{V})$\label{line:algo-1-update-current-basis}
      \If{$\scalar{w}' < \scalar{w}_{\text{min}}$}
         \State \Call{Backtrack}{$\scalar{j} + 1,\; \mymatrix{B}',\; \scalar{r}',\; \scalar{w}'$}\label{line:algo-1-recursion}
      \EndIf
   \EndFor\label{line:algo-1-search-partion-e}
\EndFunction

\State\State $\scalar{r}_{\text{init}} \gets \operatorname{rk}(\mymatrix{M})$   \Comment{$\scalar{r}_{\text{init}} = \scalar{k}$}

\For{$\scalar{j} \gets 0$ \textbf{to} $\scalar{k}' - 1$}\label{line:algo-1-parition-set-b}
   \State $\myset{C}_{\scalar{j}} \gets \{\, \addr{V}\in \myset{R} \mid \addr{V}[\scalar{j}] = 1\,\}$
   \Comment{sort $\myset{C}_{\scalar{j}}$ by increasing Hamming weight}
\EndFor\label{line:algo-1-parition-set-e}
\State\State $\mymatrix{B}_{\mathrm{best}}\gets[\,]$, \;
       $\scalar{w}_{\text{min}}\gets\infty$

\State \Call{Backtrack}{$0,\;[\,],\;\scalar{r}_{\text{init}},\;0$}\label{line:algo-1-begin-search}
\State $\mymatrix{R} \gets \mymatrix{B}_{\text{best}}$

\State\State \Return $\mymatrix{R}$
\end{algorithmic}
\end{algorithm}

\Cref{alg:rank-aware-backtrack} describes a procedure that builds a basis $\mymatrix{R}$ satisfying the above-mentioned conditions.
Firstly, the algorithm partitions the set of row masks $\myset{R}$ into totally ordered sets $\myset{C}_{\scalar{j}}=\{\addr{V} \in \myset{R} \mid \addr{V}[\scalar{j}]= 1 \}$, each sorted by Hamming weight (\Cref{line:algo-1-parition-set-b} -- \Cref{line:algo-1-parition-set-e}).
Then, the algorithm starts the search for a basis $\mymatrix{R}$ by calling the recursive function \texttt{Backtrack} (\Cref{line:algo-1-begin-search}).
\texttt{Backtrack} searches in $\myset{C}_{\scalar{j}}$ a row mask $\addr{V}$ that select the $\scalar{j}$-th row bit (\Cref{line:algo-1-search-partion-b} -- \Cref{line:algo-1-search-partion-e}). 
The function skips any row mask $\addr{V}$ that does not increase the current rank $\scalar{r}$ of $[\mymatrix{M}; \mymatrix{B}]$ (\ie the composition of $\mymatrix{M}$ and $\mymatrix{B}$ along the row-axis) (\Cref{line:algo-1-skip-dependent-mask-b} -- \Cref{line:algo-1-skip-dependent-mask-e}); otherwise, \texttt{Backtrack} adds $\addr{V}$ to the current basis $\mymatrix{B}$, updates the rank and the Hamming weight (\Cref{line:algo-1-update-current-basis}), and calls itself to explore the set $\myset{C}_{\scalar{j + 1}}$ (\Cref{line:algo-1-recursion}).
The algorithm interrupts the recursive search (i.e., it backtracks) in two occasions: when the current basis $\mymatrix{B}$ cannot reach full rank (\ie it cannot satisfy condition \textbf{C.3}) (\Cref{line:algo-1-no-full-rank-b} -- \Cref{line:algo-1-no-full-rank-e}); when the new identified basis has lower Hamming weight than the best basis found so far (\Cref{line:algo-1-found-full-basis-b} -- \Cref{line:algo-1-found-full-basis-e}). The in-order exploration of each set $\myset{C}_{\scalar{j}}$ provides a row basis $\mymatrix{R}$ whose diagonal elements are set to $1$ (\ie the algorithm satisfies condition \textbf{C.1}). The algorithm provides a final basis $\mymatrix{B}_{\mathrm{best}}$ such that $\operatorname{rk}[\mymatrix{M};\mymatrix{R}]=\operatorname{rk}(\mymatrix{M}) + \scalar{k}'$ (\ie the algorithm satisfies condition \textbf{C.3}) while minimizing the basis' Hamming weight (\ie the algorithm satisfies condition \textbf{C.2}).

Thus, from the previously retrieved bank masks and a new targeted latency analysis using these masks, we can build an addressing function for the rows.

\summarybox{Using the same methodology as for the banks in \Cref{sec:nullspace-analysis}, our previously built coarse-grained row function, and hardware-based hypotheses, we build a plausible row addressing function that explains observed latencies.}

\section{Evaluation}\label{sec:evaluation}

\subsection{Experimental setup}
\label{sec:measure}
We evaluate our methodology in a range of platforms from embedded to server-class SoCs, and for three different Instruction Set Architectures (ISA). 
For x86 targets, we use the \texttt{clflush} instruction to evict the address pairs from the cache, allowing us to be sure of measuring DRAM latencies. We use the \texttt{rdtsc} instruction for timing measurements. On ARMv8, we use \texttt{DC CIVAC} for cache eviction, and the \texttt{PMCCNTR} cycle counter for measurement. On ppc64le we flush each cache line with the \texttt{dcbf} instruction and time the access using the 64‑bit time‑base counter read by \texttt{mftb}.

\subsection{Number of Required Samples}\label{subsec:required-samples}
To ensure the practical effectiveness of our methodology, we propose a theoretical upper bound to the cardinality $\scalar{m} = \card{\myset{S}}$ (\ie the set of randomly generated address pairs) required to achieve bank mask recovery.
With $\scalar{t} = \scalar{n} - \scalar{k}$ unknown mask dimensions and latency misclassification rate $\theta \in [\,0, 1)$,
choosing $\scalar{m}$ samples such that:
\[
    \scalar{m} \geq \frac{2^k}{1 - \theta} \cdot \operatorname{log_2}(\frac{2^{\scalar{n} - \scalar{k}} - 1}{\epsilon})
\]
suffices to have \(\operatorname{rk}(\mymatrix{D}) = \scalar{t}\) with a probability greater or equal than \(1-\varepsilon\), with $\varepsilon \in [\,0, 1)$ an arbitrarily defined accepted failure probability.

We provide a similar bound on the cardinality $\scalar{m} = \card{\myset{S}_{\text{low}}}$ (\ie the set of address pairs in $\myset{S}$ exhibiting low access latency): it is sufficient to replace $\scalar{k}$ with $\scalar{k}'$ (the number of row parity masks), and $\scalar{n}$ with $\scalar{n} - \scalar{k}$ (as we have already determined the identity of $\scalar{k}$ out of $\scalar{n}$ address bits).

The complete bound's derivation is provided in \Cref{sec:sample-complexity}.

\subsection{Subsampling Technique}
\label{sec:subsampling}

\begin{figure}[t]
    \centering
    \resizebox{\linewidth}{!}{
            \begin{tikzpicture}[
  box/.style  = {draw, rounded corners, align=left,
                 inner sep=6pt, font=\normalsize},
  arr/.style  = {-{Stealth[length=3mm]}, thick},
]

\node[box] (full) {\textbf{Total pairs $D$}};

\node[box, right=2cm of full] (sub1) {%
  \textbf{Subsample $D_1$}\\[3pt]
  \\Solutions :\\[3pt]
  \colorbox{blue!20}{\ttfamily 0xA53E9000}\\
  \colorbox{purple!20}{\ttfamily 0x4C21A000}\\
  \colorbox{yellow!20}{\ttfamily 0x80000000}\\
  \colorbox{white}{\textcolor{red}{\sout{\ttfamily 0x12F00000}}}
};
\node[box, below=0.3cm of sub1] (sub2) {%
  \textbf{Subsample $D_2$}\\[3pt]
  \\Solutions :\\[3pt]
  \colorbox{blue!20}{\ttfamily 0xA53E9000}\\
  \colorbox{green!20}{\ttfamily 0x00000800}\\
  \colorbox{yellow!20}{\ttfamily 0x80000000}\\
  \colorbox{white}{\textcolor{red}{\sout{\ttfamily 0x03C40000}}}
};
\node[box, above=0.3cm of sub1] (sub3) {%
  \textbf{Subsample $D_3$}\\[3pt]
  \\Solutions :\\[3pt]
  \colorbox{purple!20}{\ttfamily 0x4C21A000}\\
  \colorbox{yellow!20}{\ttfamily 0x80000000}\\
  \colorbox{green!20}{\ttfamily 0x00000800} \\
  \colorbox{white}{\textcolor{red}{\sout{\ttfamily 0xDEADA550}}}
};

\node[box, right=2cm of sub1, yshift=-0.15cm] (vote) {%
  \textbf{Majority vote}\\[3pt]
  \colorbox{green!20}{\ttfamily 0x00000800}\\
  \colorbox{purple!20}{\ttfamily 0x4C21A000}\\
  \colorbox{yellow!20}{\ttfamily 0x80000000}\\
  \colorbox{blue!20}{\ttfamily 0xA53E9000}
};

\foreach \s in {sub1,sub2,sub3}{
   \draw[arr] (full) -- (\s);
   \draw[arr] (\s) -- (vote);
}
\end{tikzpicture}
    }
    \caption{Subsampling with majority vote. Spurious masks (red strike‑through) appear in only one subsample and are discarded; masks present in at least two subsamples survive in the final set.}
    \label{fig:subsampling}
\end{figure}
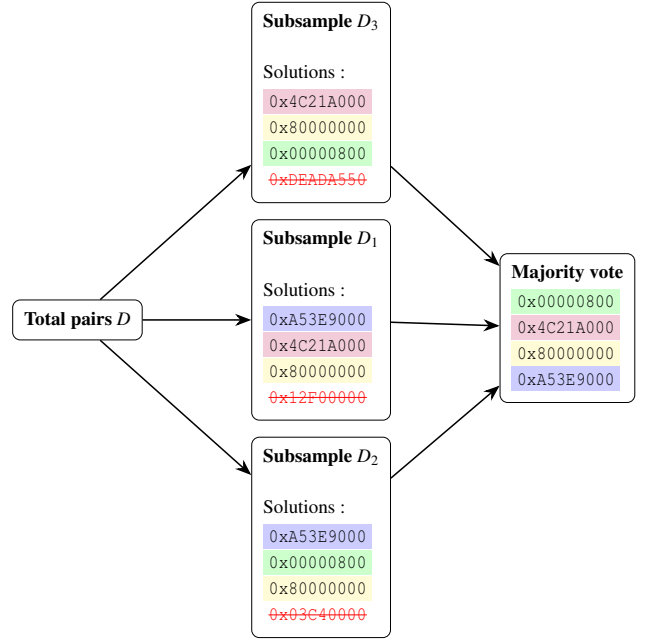

Empirical latency measurements inherently contain noise, causing occasional misclassifications of address pairs as conflicting or non-conflicting. Because such pairs do not verify the same relationships as their correctly-labeled counterparts, they increase the rank of the difference matrix $D$, thus decreasing the dimension of the nullspace and reducing the number of found masks.
To handle this noise, we introduce a subsampling technique illustrated in \Cref{fig:subsampling}:
\begin{itemize}[topsep=0pt]
    \item Instead of solving only once, we run our methodology on multiple smaller sets of pairs.
    \item Final masks are chosen by a majority vote, selecting masks that appear the most consistently.
\end{itemize}

\subsection{Results and Validation}
\begin{table*}[t]
\centering
\caption{Retrieved parity masks and address mappings for evaluated platforms.}
\rowcolors{2}{rowgray}{white}
\begin{tabular}{@{}p{5cm}>{\centering\arraybackslash\small}p{1.2cm}>{\centering\arraybackslash\small}p{4.5cm}>{\centering\arraybackslash\small}p{2cm}>{\centering\arraybackslash\small}p{3cm}@{}}
\toprule
\textbf{Platform} & \textbf{Architecture} & \textbf{SoC} & \textbf{DRAM} & \textbf{Comparison to Previous Work}\\ 
\midrule
Raspberry Pi 3B+ & ARMv8 & Broadcom BCM2837B0 & 1 GB LPDDR2 & Matches~\cite{kaur_flipping_2023}\\
Google Pixel 3a & ARMv8 & Snapdragon 670 & 4 GB LPDDR4 & New\\
Switch P4 & x86 & Intel Pentium D1517 & 8 GB DDR4 & New\\
Dell Precision Tower 5810 & x86 & Intel E5-1650 v3 & 32 GB DDR4 & New\\
Dell Precision Tower 7875 & x86 & AMD Threadripper PRO 7955WX & 64 GB DDR5 & New\\
Dell PowerEdge R630       & x86   & Intel Xeon E5‑2630 v3        & 128 GB DDR4 & Matches~\cite{pessl_drama_2016}\\
HPE Proliant DL360 Gen10+ & x86 & Intel Xeon Silver 4314 & 256 GB DDR4 & New\\
ThinkSystem SR630 V2 & x86 & Intel Xeon Gold 5318Y & 256 GB DDR4 & New\\
Nvidia DGX-1 & x86 & Intel Xeon E5-2698 v4 & 512 GB DDR4 & New\\
IBM PowerNV S822LC & ppc64le & IBM POWER8NVL 1.0 & 128 GB DDR4 & New\\
\bottomrule
\end{tabular}
\label{tab:retrieved-masks}
\end{table*}


To validate our methodology and the implemented code, we first use synthetic data generated from functions retrieved by previous works~\cite{kogler_half-double_2022} to check if \knocknock is able to give back the same mappings found by these works. 
In a second step, we use real-world datasets obtained by running \knocknock on target platforms as shown in \Cref{tab:retrieved-masks}.

In a first step, for targets with known mappings that we did not have access to, we predict whether a conflict would happen or not between two randomly generated addresses using the documented functions. Then, we input this artificial data into our algorithm to check its ability to retrieve the functions and compare them back with the mappings found in the original work~\cite{kogler_half-double_2022}.
We run experiments using a synthetic dataset built from up to 100,000 randomly generated addresses, and we find a 100\% accuracy in the mapping functions found.
With this method, we could also test for the tolerance to noise. We did so by modifying the dataset with up to 5\% of misclassified timings.
The results in this case, still giving a 100\% accuracy of recovering, showed how the methodology can tolerate measurement noise.

Secondly, we evaluate the accuracy of our masks by measuring their ability to correctly predict high and low-latency address pairs. These are generated through an additional validation step using 10,000 new latency measurements.
For this real-world data, we leverage the privileged \texttt{/proc/pagemap} interface to translate virtual addresses to physical addresses, and measure access latencies using high-resolution timers.
We discuss how this requirement does not nullify the threat model in \Cref{sec:discussion}.
When available, the recovered masks are validated against known hardware configurations or previous works.
For the E5-2630 v3, while DRAMA~\cite{pessl_drama_2016} documents found masks on their 64GB setup, we could not reproduce the result using their source code~\cite {dramagh} on our 128GB platform.
This is probably due to the exponential search phase, reaching its timeout after a few hours, before fully achieving parity-mask recovery.
\knocknock allowed us to retrieve similar masks to those documented in the original article in just a few minutes.
The list of platforms and their respective results is summarized in \Cref {tab:retrieved-masks} and \Cref{tab:retrieved-masks-values}.
We find functions for previously undocumented targets, across different architectures and use cases, ranging from embedded to server-class hardware.
We compare our functions with previous works~\cite{pessl_drama_2016, kaur_flipping_2023} and find similar functions when the target matches.
We use standard classification metrics like precision ($TP/{TP+FP}$) and recall ($TP/{TP+FN}$) to evaluate our results, where $TP$, $FP$, $TN$ and $FN$ are defined as follows:
\begin{description}[noitemsep, topsep=0pt, leftmargin=1.5em]
    \item $TP$: \textbf{True positive} - Predicted and real conflict;
    \item $FP$: \textbf{False positive} - Predicted conflict, real non-conflict;
    \item $TN$: \textbf{True negative} - Predicted and real non conflict;
    \item $FN$: \textbf{False negative} - Predicted non-conflict, real conflict.
\end{description}
For each target, we evaluate the recall and the precision and find that they are both greater than 99\%, meaning that the found masks correctly explain the observed conflicts.

\summarybox{Our method works with $>99\%$ recall and precision, even in noisy environments. The results match previous reverse-engineering works, with a faster search phase and a black-box approach.}

\section{Discussion}\label{sec:discussion}

\textbf{Closed-Page Policy:}
We tested some platforms, notably the Raspberry Pi 4 4GB and the Nvidia Jetson Nano, for which we were unable to retrieve any part of the mapping.
The latency measurements showed a single distribution, indicating a closed-page policy.
A closed-page policy means that the memory controller closes the row after each access~\cite{blackmore_quantitative_2000}, which prevents us from differentiating between row hits and row conflicts. Because the row buffer is always closed after use, each memory access has to be served from the bank, eliminating the lower latency peak seen in \Cref{fig:histogram}.

\textbf{Privileged access to the pagemap:}
While we assumed privileged access to the \texttt{/proc/pagemap} interface, which is a common threat model~\cite{wang_dramdig_2020, pessl_drama_2016}, we consider that this does not invalidate the attack vector that this work makes possible.
Indeed, for tested devices, the mapping was consistent across different devices of the same model.
Therefore, an attacker owning the same device as their victim could use the knowledge obtained from characterising their device, and use it later on from a malicious process on the victim's device, even if it only has access to the virtual address~\cite{bechtel_memory-aware_2022, kogler_half-double_2022}.
The labeling methodology described in Sudoku~\cite{wi_sudoku_2025} could be used as a post-processing step on top of \knocknock to improve human readability, although it does not bring any improvement in the considered use case of attacks building on known DRAM address mappings.

\textbf{Ground truth:}
Our attack being purely based on software, we consider the translation from physical address to DRAM rows.
This addressing relies on several different mappings, such as the memory controller's mapping and the DRAM's internal addressing.
Our method cannot disentangle these mappings; instead, it treats the entire translation pipeline as a single black box.
While this remains sufficient to mount attacks on the DRAM, such as Rowhammer, this approach may present limitations.
For instance, modifying the CPU configuration may affect the memory controller map only, and not the DRAM internal mapping.
This approach remains the state-of-the-art standard for software-based physical-to-DRAM addressing~\cite{wang_dramdig_2020, helm_reliable_2020, wi_sudoku_2025, heckel2023reverse,jattke_zenhammer_2024}, where we improve previous methods with speed, reliability, and portability.
Only a few works~\cite{pessl_drama_2016,CojocarKPTSWM20} leveraged physical probing to get the ground truth necessary to reverse the memory controller mapping.
We believe combining our method with physical probing could enable separating the memory controller and DRAM internal mappings, providing a more complete understanding of the translation process, potentially resulting in a finer-grained control of the DRAM.
Such physical probing could also confirm the efficiency of our method by ruling out other sources of contention when using the row-buffer side channel.

\textbf{Hypotheses:}
Another limitation of \knocknock lies in the hypotheses of \Cref{sec:plausible-basis}.
While we only need them to reverse physical-to-row mappings, which is one of the new contributions of \knocknock, we acknowledge that they still reduce the portability of our method.
Without these, the indexing of rows might be false, resulting in an incorrect mapping.
However, we consider these hypotheses realistic, as they were verified on all tested systems in the evaluation and compared with previous work when available~\cite{pessl_drama_2016, kogler_half-double_2022}.

\textbf{Applications on countermeasures}
Due to its automated and fast approach, \knocknock can be used to defeat mitigations based on DRAM-address randomization~\cite{MeadowsEC20}.
By just running \knocknock after each SoC boot, effectively reversing the randomization, an attacker could still implement attacks requiring physical-to-DRAM-addressing functions in a reasonable time.
We leave the evaluation of the impact of \knocknock on existing countermeasures to future work.

\section{Conclusion}
\label{sec:conclusion}
In this work, we have presented \emph{Knock-Knock}, a black-box and platform-agnostic methodology that formalizes physical-to-DRAM address-mapping reverse engineering.
To achieve so, we have developed an analytical and provable methodology that bounds the complexity of the search space of possible DRAM addressing functions.
The key contribution of \knocknock is the formalization of physical-to-DRAM mapping reverse engineering, which enables retrieving functions with only timing measurements and elementary linear algebra.
This formalization allows for improved noise resilience, which we tackled during the analytical phase of the methodology, compared to previous works that have tried to improve mislabeling during the data generation and reversing phase.
We have validated our method on 10 machines, spanning from embedded SoCs to server-grade clusters, and achieved a 99\% recall and precision rate on all targets, running in only a few minutes on systems with up to 512GB of DRAM. 
\knocknock paves the way to more extensive, precise, and generic studies on microarchitectural security of memory systems across different system classes and architectures.
To that end, we publish our code and data in a public repository\footnote{https://github.com/antpln/Knock-Knock}.

    \ifAnon
    \else
\section*{Acknowledgments}
Experiments presented in this paper were done on Grid'5000 testbed, supported by a scientific interest group hosted by Inria and including CNRS, RENATER and several Universities as well as other organizations (see \url{https://www.grid5000.fr}). 
This work is funded by the French Agence Nationale de la Recherche (ANR) Young Researchers (JCJC) program, under grant number ANR-21-CE39-0018 (project ATTILA).
    \fi


\bibliographystyle{plain}
\bibliography{references}

\begin{thebibliography}{10}

\bibitem{BaiHCL25}
Yang Bai, Yizhi Huang, Si~Chen, and Renfa Li.
\newblock {PaLLOC: Pairwise-based low-latency online coordinated resource manager of last-level cache and memory bandwidth on multicore systems}.
\newblock {\em Journal of Systems Architecture}, 164:103427, 2025.

\bibitem{bechtel_memory-aware_2022}
Michael~Garrett Bechtel and Heechul Yun.
\newblock {Memory-Aware Denial-of-Service Attacks on Shared Cache in Multicore Real-Time Systems}.
\newblock {\em {IEEE} Transactions on Computers}, 71(9):2351--2357, 2022.

\bibitem{blackmore_quantitative_2000}
Matthew Blackmore.
\newblock {A Quantitative Analysis of Memory Controller Page Policies}.
\newblock Master's thesis, Portland State University, 2013.

\bibitem{Bromwich_1991}
Thomas John~I’Anson Bromwich.
\newblock {\em An introduction to the theory of infinite series}.
\newblock AMS Chelsea Publishing, 1991.

\bibitem{CojocarKPTSWM20}
Lucian Cojocar, Jeremie~S. Kim, Minesh Patel, Lillian Tsai, Stefan Saroiu, Alec Wolman, and Onur Mutlu.
\newblock {Are We Susceptible to Rowhammer? An End-to-End Methodology for Cloud Providers}.
\newblock In {\em {2020 IEEE Symposium on Security and Privacy (SP)}}, pages 712--728. {IEEE}, 2020.

\bibitem{Derya2024}
Kemal Derya, M.~Caner Tol, and Berk Sunar.
\newblock {FAULT+PROBE: A Generic Rowhammer-based Bit Recovery Attack}.
\newblock In {\em Proceedings of the 20th ACM Asia Conference on Computer and Communications Security}, ASIA CCS '25, page 1219–1234. ACM, 2025.

\bibitem{frigo_trrespass_2020}
Pietro Frigo, Emanuele Vannacci, Hasan Hassan, Victor van~der Veen, Onur Mutlu, Cristiano Giuffrida, Herbert Bos, and Kaveh Razavi.
\newblock {TRRespass: Exploiting the Many Sides of Target Row Refresh}.
\newblock In {\em 2020 IEEE Symposium on Security and Privacy (SP)}, pages 747--762. {IEEE}, 2020.

\bibitem{DBLP:conf/sp/GerlachSFS24}
Lukas Gerlach, Simon Schwarz, Nicolas Faro{\ss}, and Michael Schwarz.
\newblock {Efficient and Generic Microarchitectural Hash-Function Recovery}.
\newblock In {\em 2024 {IEEE} Symposium on Security and Privacy ({SP})}, pages 3661--3678. {IEEE}, 2024.

\bibitem{heckel2023reverse}
Martin Heckel and Florian Adamsky.
\newblock {Reverse-Engineering Bank Addressing Functions on AMD CPUs}.
\newblock In {\em 3rd Workshop on DRAM Security (DRAMSec)}, 2023.

\bibitem{helm_reliable_2020}
Christian Helm, Soramichi Akiyama, and Kenjiro Taura.
\newblock {Reliable Reverse Engineering of Intel {DRAM} Addressing Using Performance Counters}.
\newblock In {\em 2020 28th International Symposium on Modeling, Analysis, and Simulation of Computer and Telecommunication Systems (MASCOTS)}, pages 1--8. {IEEE}, 2020.

\bibitem{DBLP:conf/ccs/HofmannFKV24}
Jana Hofmann, C{\'{e}}dric Fournet, Boris K{\"{o}}pf, and Stavros Volos.
\newblock {Gaussian Elimination of Side-Channels: Linear Algebra for Memory Coloring}.
\newblock In {\em Proceedings of the 2024 on ACM SIGSAC Conference on Computer and Communications Security}, CCS '24, pages 2799--2813. {ACM}, 2024.

\bibitem{dramagh}
isec{-}tugraz.
\newblock {DRAMA - Source code repository}.
\newblock \url{https://github.com/isec-tugraz/drama}, June 2016.
\newblock Commit used : c5c8347.

\bibitem{jattke_zenhammer_2024}
Patrick Jattke, Max Wipfli, Flavien Solt, Michele Marazzi, Matej B{\"{o}}lcskei, and Kaveh Razavi.
\newblock {ZenHammer: Rowhammer Attacks on {AMD} Zen-based Platforms}.
\newblock In {\em 33rd USENIX Security Symposium (USENIX Security 24)}, pages 1615--1633, Philadelphia, PA, 2024. {USENIX} Association.

\bibitem{lpddr4standard}
JC-42.
\newblock {Low Power Double Data Rate 4 ({LPDDR}4) {\textbar} {JEDEC}}.

\bibitem{ddr4standard}
JC-42.3C.
\newblock {DDR}4 {SDRAM} {STANDARD} {\textbar} {JEDEC}.

\bibitem{jung_reverse_2016}
Matthias Jung, Carl~C. Rheinländer, Christian Weis, and Norbert Wehn.
\newblock {Reverse Engineering of {DRAMs}: Row Hammer with Crosshair}.
\newblock In {\em Proceedings of the Second International Symposium on Memory Systems}, {MEMSYS} '16, pages 471--476. ACM, 2016.

\bibitem{katznelson_terse_2008}
Yitzhak Katznelson and Yonatan~R. Katznelson.
\newblock {\em {A (Terse) Introduction to Linear Algebra}}.
\newblock American Mathematical Soc., 2008.

\bibitem{kaur_flipping_2023}
Anandpreet Kaur, Pravin Srivastav, and Bibhas Ghoshal.
\newblock {Flipping Bits Like a Pro: Precise Rowhammering on Embedded Devices}.
\newblock {\em {IEEE} Embedded Systems Letters}, 15(4):218--221, 2023.

\bibitem{kim_flipping_2014}
Yoongu Kim, Ross Daly, Jeremie~S. Kim, Chris Fallin, Ji{-}Hye Lee, Donghyuk Lee, Chris Wilkerson, Konrad Lai, and Onur Mutlu.
\newblock Flipping bits in memory without accessing them: An experimental study of {DRAM} disturbance errors.
\newblock In {\em 2014 ACM/IEEE 41st International Symposium on Computer Architecture (ISCA)}, pages 361--372. {IEEE} Computer Society, 2014.

\bibitem{kogler_half-double_2022}
Andreas Kogler, Jonas Juffinger, Salman Qazi, Yoongu Kim, Moritz Lipp, Nicolas Boichat, Eric Shiu, Mattias Nissler, and Daniel Gruss.
\newblock {Half-Double: Hammering From the Next Row Over}.
\newblock In {\em 31st USENIX Security Symposium (USENIX Security 22)}, pages 3807--3824. {USENIX} Association, 2022.

\bibitem{ZebRAM}
Radhesh~Krishnan Konoth, Marco Oliverio, Andrei Tatar, Dennis Andriesse, Herbert Bos, Cristiano Giuffrida, and Kaveh Razavi.
\newblock {ZebRAM: Comprehensive and Compatible Software Protection Against Rowhammer Attacks}.
\newblock In {\em 13th USENIX Symposium on Operating Systems Design and Implementation (OSDI 18)}, pages 697--710. {USENIX} Association, 2018.

\bibitem{kwong_rambleed_2020}
Andrew Kwong, Daniel Genkin, Daniel Gruss, and Yuval Yarom.
\newblock {RAMBleed: Reading Bits in Memory Without Accessing Them}.
\newblock In {\em 2020 {IEEE} Symposium on Security and Privacy ({SP})}, pages 695--711. {IEEE}, 2020.

\bibitem{marazzi2024risc}
Michele Marazzi and Kaveh Razavi.
\newblock {RISC-H: Rowhammer Attacks on RISC-V}.
\newblock In {\em 4th Workshop on DRAM Security (DRAMSec)}, 2024.

\bibitem{MeadowsEC20}
Brett Meadows, Nathan Edwards, and Sang{-}Yoon Chang.
\newblock {On-Chip Randomization for Memory Protection Against Hardware Supply Chain Attacks to {DRAM}}.
\newblock In {\em 2020 IEEE Security and Privacy Workshops (SPW)}, pages 171--180. {IEEE}, 2020.

\bibitem{PanGM16}
Xing Pan, Yasaswini~Jyothi Gownivaripalli, and Frank Mueller.
\newblock {TintMalloc: Reducing Memory Access Divergence via Controller-Aware Coloring}.
\newblock In {\em 2016 IEEE International Parallel and Distributed Processing Symposium (IPDPS)}, pages 363--372. {IEEE} Computer Society, 2016.

\bibitem{pessl_drama_2016}
Peter Pessl, Daniel Gruss, Cl{\'{e}}mentine Maurice, Michael Schwarz, and Stefan Mangard.
\newblock {{DRAMA:} Exploiting {DRAM} Addressing for Cross-CPU Attacks}.
\newblock In {\em 25th USENIX Security Symposium (USENIX Security 16)}, pages 565--581. {USENIX} Association, 2016.

\bibitem{RakinCYF22}
Adnan~Siraj Rakin, Md~Hafizul~Islam Chowdhuryy, Fan Yao, and Deliang Fan.
\newblock {DeepSteal: Advanced Model Extractions Leveraging Efficient Weight Stealing in Memories}.
\newblock In {\em {2022 IEEE Symposium on Security and Privacy (SP)}}, pages 1157--1174. {IEEE}, 2022.

\bibitem{schwarz_malware_2019}
Michael Schwarz, Samuel Weiser, Daniel Gruss, Cl{\'{e}}mentine Maurice, and Stefan Mangard.
\newblock {Malware Guard Extension: Using {SGX} to Conceal Cache Attacks}.
\newblock In {\em Detection of Intrusions and Malware, and Vulnerability Assessment - 14th International Conference, {DIMVA} 2017}, volume 10327 of {\em Lecture Notes in Computer Science}, pages 3--24. Springer, 2017.

\bibitem{DBLP:journals/tc/VandierendonckB05}
Hans Vandierendonck and Koenraad~De Bosschere.
\newblock {XOR-Based Hash Functions}.
\newblock {\em {IEEE Transactions on Computers}}, 54(7):800--812, 2005.

\bibitem{wang_dramdig_2020}
Minghua Wang, Zhi Zhang, Yueqiang Cheng, and Surya Nepal.
\newblock {DRAMDig: {A} Knowledge-assisted Tool to Uncover {DRAM} Address Mapping}.
\newblock In {\em 2020 57th ACM/IEEE Design Automation Conference (DAC)}, pages 1--6. {IEEE}, 2020.

\bibitem{wang2020figaro}
Yaohua Wang, Lois Orosa, Xiangjun Peng, Yang Guo, Saugata Ghose, Minesh Patel, Jeremie~S. Kim, Juan~Gómez Luna, Mohammad Sadrosadati, Nika~Mansouri Ghiasi, and Onur Mutlu.
\newblock {FIGARO: Improving System Performance via Fine-Grained In-DRAM Data Relocation and Caching}.
\newblock In {\em 2020 53rd Annual IEEE/ACM International Symposium on Microarchitecture (MICRO)}, pages 313--328, 2020.

\bibitem{wi_sudoku_2025}
Minbok Wi, Seungmin Baek, Seonyong Park, Mattan Erez, and Jung~Ho Ahn.
\newblock {Sudoku: Decomposing DRAM Address Mapping into Component Functions}.
\newblock In {\em 5th Workshop on DRAM Security (DRAMSec)}, 2025.

\bibitem{WiPKKKLA23}
Minbok Wi, Jaehyun Park, Seoyoung Ko, Michael~Jaemin Kim, Nam~Sung Kim, Eojin Lee, and Jung~Ho Ahn.
\newblock {{SHADOW:} Preventing Row Hammer in {DRAM} with Intra-Subarray Row Shuffling}.
\newblock In {\em {2023 IEEE International Symposium on High-Performance Computer Architecture (HPCA)}}, pages 333--346. {IEEE}, 2023.

\bibitem{yoon2011row}
HanBin Yoon, Justin Meza, Rachata Ausavarungnirun, Rachael Harding, and Onur Mutlu.
\newblock {Row Buffer Locality-Aware Data Placement in Hybrid Memories}.
\newblock {\em SAFARI Technical Report}, 5, 2011.

\end{thebibliography}

\clearpage

\appendix
\section{Appendix}
\begin{table*}[!t]
\centering
\caption{Retrieved parity masks and address mappings for evaluated platforms. [x,y] indicate that the masks have bits x and y set, while for masks with multiple bits we describe them using their hexadecimal representations.}
\rowcolors{2}{rowgray}{white}
\resizebox{\linewidth}{!}{%
\begin{tabular}{@{}p{4cm}>{\centering\arraybackslash}p{16cm}@{}}
\toprule
\textbf{Platform} & \textbf{Retrieved Bank/Channel Masks}\\ 
\midrule
Raspberry Pi 3B+ & \texttt{[13], [14], [15]} \\
Google Pixel 3a & \texttt{0x274e9000}, \texttt{0x69d3a000}, \texttt{0x53a74000}, \texttt{0x80000000}\\
Switch P4 & \texttt{[6,20], [17, 21], [18,22], [19, 23], [32, 33]} \\
Dell Precision Tower 5810 & \texttt{0x8000}, \texttt{0x100000000}, \texttt{0x200000000}, \texttt{0x400000000}, \texttt{0x800040}, \texttt{0x1100000}, \texttt{0x2200000}, \texttt{0x4400000}, \texttt{0x55080}, \texttt{0x88a2100} \\
Dell Precision Tower 7875 & \texttt{0x84201000}, \texttt{0x40214100}, \texttt{0x188400200}, \texttt{0x1421002000}, \texttt{0x310800400}, \texttt{0x1842100800}, \texttt{0xff80000}, \texttt{0xd6f700440}\\
Dell PowerEdge R630 & \texttt{0x800040}, \texttt{0xa00000000}, \texttt{0xc00000000}, \texttt{0x3000000000}, \texttt{0x4408000}, \texttt{0x2820000000}, \texttt{0x5500000}, \texttt{0x6600000}, \texttt{0x88a2100}, \texttt{0x4455080}\\
HPE Proliant DL360 Gen10+ & \texttt{[15], [35], [36], [37], [6,23], [20,24], [21,25], [22, 26], 0x4004100, 0x6024800}\\
Nvidia DGX-1 & \texttt{[37], [38], [16], [15], [21,25], [6, 24], [7, 17], [23, 27], [22, 26], [8, 12, 14, 18, 20, 24]}\\
ThinkSystem SR630 V2 & \texttt{[16], [35], [36],  [37], [6, 24], [21, 25], [22, 26], [23, 27], [8, 14, 26], [9, 15, 27], [11, 14, 17, 25, 26]}\\
IBM PowerNV S822LC & \texttt{[7],[8],[9],[10],[11],[12],[13],[14],[15],[32,34],[33,34]}
\\
\bottomrule
\end{tabular}
}
\label{tab:retrieved-masks-values}
\end{table*}

\subsection{Sample Complexity for Recovering the Bank/Channel Masks}
\label{sec:sample-complexity}
For our methodology, it is important to randomly generate a sufficient number $\scalar{m}$ of $\scalar{n}$-bit address pairs to satisfy \Cref{eq:rev-eng-proof-nullspace-equiv} (\Cref{sec:nullspace-analysis}).

In practice, we have to bound $s$ such that:
\begin{equation}
    P[\operatorname{rk}(\mymatrix{D}) = \scalar{t}] \geq 1 - \epsilon,
    \label{eq:appendix-probability-linear-independence}
\end{equation}

 where $\mymatrix{D}$ is the difference matrix of size $m \times n$, $t = n - k$ is the targeted rank for $\mymatrix{D}$, $\scalar{k}$ is the dimension of nullspace($\mymatrix{D}$), and $\epsilon \in [\,0, 1)$ is the arbitrarily defined accepted failure probability.
We define $\theta \in [\,0, 1)$ as the proportion of misclassified pairs, \ie the proportion of address pairs that are not conflicts but are classified as such.
We do not consider $\epsilon = 1$ and $\theta = 1$ as, in such cases, it is not possible to recover the bank and channel masks.

Then, we define $\scalar{m}_{c}$ as the number of address pairs classified as conflicts.
Therefore, for a required $\scalar{m}_{c}$ conflicting pairs, we need to randomly generate $\scalar{m}$ address pairs such that:

\begin{equation}
    \scalar{m} =\frac{2^{\scalar{k}}}{1 - \theta} \cdot \scalar{m}_{c}.\label{eq:appendix-number-samples}
\end{equation}

To simplify the computation, we calculate a lower bound for this probability using $\mymatrix{D}_{t} \in \{0, 1\}^{\scalar{m}_{c} \times \scalar{t}}$, which only contains the $\scalar{t}$ first columns of $\mymatrix{D}$.

By construction :
\[
P[\operatorname{rk}(\mymatrix{D}) = \scalar{t}] \geq P[\operatorname{rk}(\mymatrix{D}_{t}) = \scalar{t}].
\]

We denote $d_{j}$ as the $j$-th column of $\mymatrix{D}_{\scalar{t}}$. We remark that all $\mymatrix{D}_{\scalar{t}}$'s columns are randomly drawn with replacement from a uniform distribution over $\{0, 1\}^{\scalar{t}}$.

Let us assume that we have already found $\scalar{j}$ linearly independent columns.
These columns span a set of vectors of size $2^{\scalar{j}}$.
Therefore, the probability that column $d_{\scalar{j} + 1}$ is linearly independent from the already chosen columns is:
\[
    P[d_{\scalar{j} + 1} \text{ is linearly independent}] = 1 - \frac{2^{\scalar{j}}}{2^{\scalar{m}_{c}}} = 1 - 2^{\scalar{j} - {\scalar{m}_{c}}}.
\]
Then, the probability of randomly drawing $t$ linearly independent rows is:
\[
    P[\operatorname{rk}(\mymatrix{D}_{\scalar{t}}) = \scalar{t}] = \prod_{\scalar{j}  = 0}^{\scalar{t} - 1} \left(1 - 2^{\scalar{j} - {\scalar{m}_{c}}}\right).
\]
According to the Weierstrass' product inequality~\cite{Bromwich_1991}:

\begin{equation}
    P[\operatorname{rk}(\mymatrix{D}_{\scalar{t}}) = \scalar{t}] \geq 1 - 2^{-{\scalar{m}_{c}}} \cdot\sum_{\scalar{j} = 0}^{\scalar{t} - 1}\left(2^{\scalar{j}} - 1\right) = 1 - \frac{2^{\scalar{t}} - 1}{2^{\scalar{m}_{c}}}
    \label{eq:appendix:weirstrass-application}
\end{equation}

Combining \Cref{eq:appendix-probability-linear-independence} and \Cref{eq:appendix:weirstrass-application}, we have that:
\begin{equation}
    \epsilon = \frac{2^\scalar{t} - 1}{2^{\scalar{m}_{c}}}.
    \label{eq:appendix-epsilon-expression}
\end{equation}

By applying \Cref{eq:appendix-number-samples} to \Cref{eq:appendix-epsilon-expression}, we find that \Cref{eq:appendix-probability-linear-independence} is satisfied for:
\\
\[
\boxed{
    \scalar{m} \geq \frac{2^{\scalar{k}}}{1 - \theta} \cdot \operatorname{log_2}(\frac{2^{\scalar{n} - \scalar{k}} - 1}{\epsilon}).
}
\]\\
As an example, for $\scalar{n} = 32$ bit addresses (\ie 4 GB Memory), $\scalar{k} = 4$ (\ie 8 banks and 2 channels), $\theta = 0.05$ (\ie 5\% misclassifications), and $\epsilon = 0.01$ we would need $\scalar{m} \geq 584$
randomly generated address pairs to guarantee the identification of parity masks for banks and channels.

To guarantee the recovery of the row parity masks, we follow the same reasoning to provide a bound similar to the one we have derived for bank parity masks:
\\
\[
\boxed{
    \scalar{m}' \geq \frac{2^{\scalar{k}'}}{1 - \theta} \cdot \operatorname{log_2}(\frac{2^{\scalar{n} - \scalar{k} - \scalar{k}'} - 1}{\epsilon}).
}
\]\\

Using the same parameters defined in the previous example, and for $\scalar{k}' = 4$ row bits, we would need $\scalar{m}' \geq 517$ address pairs exhibiting low access latency.

\end{document}